\documentclass[runningheads]{llncs}

\usepackage[mobile]{eccv}

\usepackage{eccvabbrv}

\usepackage{graphicx}
\usepackage{booktabs}

\usepackage{multirow}

\usepackage[nomargin,inline,final]{fixme}
\fxsetup{theme=color,mode=multiuser}
\FXRegisterAuthor{ra}{ara}{\color{blue}roberto}
\FXRegisterAuthor{lr}{lre}{\color{orange}lucas}

\usepackage[accsupp]{axessibility}  %

\usepackage{hyperref}

\usepackage{orcidlink}

\begin{document}

\title{Region-Adaptive Generative Compression with Spatially Varying Diffusion Models} 

\titlerunning{Region-Adaptive Compression with Diffusion}

\author{
Lucas Relic\inst{1,2}%
\and Roberto Azevedo\inst{2}%
\and Yang Zhang\inst{2}
\and Stephan Mandt\inst{3}
\and Markus Gross\inst{1,2}%
\and Christopher Schroers\inst{2}%
}

\authorrunning{L.~Relic et al.}

\institute{ETH Z\"urich; Z\"urich, Switzerland \\
\and
Disney Research\textbar Studios; Z\"urich, Switzerland \\
\and
UC Irvine; California, USA
}

\maketitle

\begin{abstract}
Generative image codecs aim to optimize perceptual quality, producing realistic and detailed reconstructions.
However, they often overlook a key property of human vision: our tendency to focus on particular aspects of a visual scene (e.g., salient objects) while giving less importance to other regions.
An ideal perceptual codec should be able to exploit this property by allocating more representational capacity to perceptually important areas.
To this end, we propose a region-adaptive diffusion-based image codec that supports non-uniform bit allocation within an image.
We design a novel spatially varying diffusion model capable of denoising varying amounts of noise per pixel according to arbitrary importance maps.
We further identify that these maps can serve as effective priors on the latent representation, and integrate them into our entropy model, improving rate-distortion performance.
Built on these contributions, our spatially-adaptive diffusion-based codec outperforms state-of-the-art ROI-controllable baselines in both full-image and ROI-masked perceptual quality. 
\end{abstract}
    
\section{Introduction}
\label{sec:intro}

\begin{figure}
    \centering
    \includegraphics[width=0.95\linewidth]{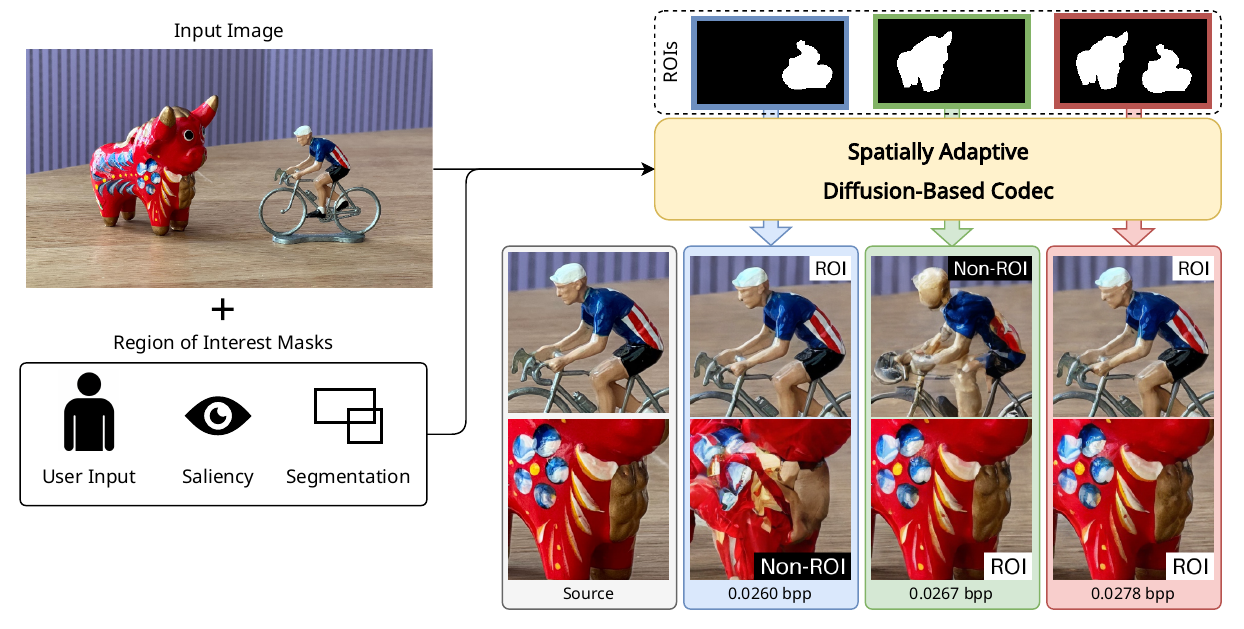}
    \caption{High-level illustration of our region-aware generative image compression framework. Based on the region of interest, our method adaptively allocates bits for different areas of the image. Regions of interest are devoted more representational capacity, and thus are more accurately reconstructed. Content in less important areas is expectedly more generated, but still seamlessly integrates into the scene while looking as realistic as possible.}
    \label{fig:teaser}
\end{figure}

Recently, the integration of generative modeling into learned image compression frameworks has transformed the field, enabling realistic reconstructions at extremely low bitrates.
Focusing on the end-user perceptual experience, unlike traditional codecs, generative compression methods emphasize visual realism rather than strict pixel fidelity,
which closely aligns with human visual preferences~\cite{mentzer2020HighFidelitya}.
However, current methods overlook an important aspect of the human vision system: \emph{attentional selection}~\cite{goldstein2014Sensation}.

The human visual system is inherently spatially adaptive;
attention is naturally drawn to salient objects and regions, whereas peripheral or background areas can tolerate higher levels of distortion without noticeable degradation~\cite{rozenholtz2011What, kanwisher2000Visual}.
Consequently, codecs that emphasize perceptual quality should exploit this property by allocating more representational capacity to perceptually important regions.
Despite this potential, explicit spatial control remains largely underexplored in generative compression.

Classical and non-generative learned image codecs have attempted to leverage spatial importance through region-of-interest (ROI)–based bit allocation.
Although this offers strong spatial control, allocating fewer bits outside the ROI in these methods often results in extremely poor performance in those areas~(seen in \cref{fig:roi_qualitative}, C-ROI and TV-ROI).
Such artifacts are further amplified by the tendency of non-generative methods to produce high-fidelity, but not necessarily realistic, reconstructions, which contrasts with the motivation to improve the end users' perceptual experience.

Focused on the above research gap, we propose a novel region-adaptive diffusion-based image codec~(\cref{fig:teaser}).
Our approach introduces a spatially adaptive diffusion model with a uniform noise distribution, enabling compression with dynamic bit allocation according to local perceptual importance.
We additionally build on the intuition that more compressed regions need more denoising~(i.e., more generation) and explicitly draw a connection between this visual importance map and diffusion timestep.
Correspondingly, we introduce a timestep-conditioned entropy model leveraging such importance maps as prior information during coding and demonstrate it improves performance.
By combining the strengths of diffusion models with the principle of spatial adaptivity, our approach establishes a new framework for region-adaptive generative compression.
On the one hand, compared to classical and non-generative compression methods, our approach consistently supports the generation of realistic content, in both ROI and non-ROIs.
On the other hand, it improves on previous diffusion-based generative compression methods by explicitly supporting spatially adaptive bitrate control, which, as shown by subjective user experiments, better aligns with human preference. 

In summary, we:
\begin{itemize}
    \item propose a spatially adaptive uniform diffusion model that denoises spatially varying amounts of noise at each timestep, and introduce a corresponding spatially adaptive sampling algorithm.
    \item introduce a novel timestep conditioned entropy model that utilizes the visual importance map as an additional prior on the latent distribution to improve compression efficiency.
    \item develop, based on the individual components above, a region-adaptive diffusion-based image codec that outperforms the previous state-of-the-art spatially adaptive methods in global and ROI-masked perceptual quality.
\end{itemize}

\vspace{-1.2em}
\section{Related Work}
\label{sec:rel_work}
\vspace{-.5em}
\subsection{ROI-based Compression}
Motivated by attentional selection of human vision, ROI-based codecs allocate bits non-uniformly across an image, dedicating more representational capacity to perceptually or semantically important regions while allowing higher compression in less critical areas.
Classical codecs such as JPEG2000~\cite{taubman2002JPEG2000a} can intuitively implement ROI functionality by applying weighting schemes to block-based quantization, but these often introduce visible discontinuities and unnatural transitions between regions.
Feng~\etal~\cite{feng2023Semantically} extended this idea to learned methods by proposing variably-quantized bit groups in a VAE-based image codec.
Other works implement region-adaptive bit allocation via spatially-adaptive feature scaling~\cite{song2021VariableRate, jin2025Customizable} or pruning~\cite{li2021Learning, xu2025Decouplea}, or by optimizing for weighted losses based on learned importance maps~\cite{patel2021Saliency, luo2024SuperHighFidelity}.
While these methods improve rate–distortion performance in salient areas, most rely on pixelwise distortion metrics~(e.g., MSE or MS-SSIM) that do not align well with human perceptual quality.
Furthermore, the lack of a generative prior often results in significantly degraded image quality in background regions.

Different from the above, by focusing on diffusion-based generative compression, our framework consistently generates realistic content in both ROI and non-ROI areas, seamlessly integrating them into a coherent scene.

\vspace{-1em}
\subsection{Diffusion-based Compression}
\label{sec:related_work_diffusion_compression}
Diffusion-based image codecs~\cite{theis2022Lossy, yang2023Lossy, relic2024Lossya} exploit the powerful generative priors of diffusion models to produce highly realistic and perceptually detailed reconstructions.
These approaches generally follow two main paradigms: conditional generation and diffusion dequantization.

Most previous works adopt the conditional generation paradigm, where compression is formulated as a conditional image synthesis task~\cite{yang2023Lossy, careil2024image, bachard2024CoCliCo, li2025Extreme}.
Compact signals derived from the source image are transmitted to the decoder and used to guide the reconstruction process via a conditional diffusion model.
These conditioning signals can take a variety of structural and semantic forms, for example, learned latent embeddings~\cite{yang2023Lossy, li2025Extreme, xia2024DiffPC}, hyperlatents~\cite{careil2024image, xu2025Decouplea}, text embeddings~\cite{bachard2024CoCliCo}, or even images compressed using non-generative codecs~\cite{hoogeboom2023HighFidelity, ghouse2023Residual}.

In contrast, the diffusion dequantization paradigm builds on the observation that quantization error resembles additive noise~\cite{balle2017Endtoend, balle2018Variationala}.
Codecs in this category incorporate quantization directly into the diffusion forward process, using learned~\cite{relic2024Lossya} or universal quantization~\cite{relic2025Bridging, yang2024Progressive} to add noise to the compressed image latents.
The reverse diffusion process is then initialized with this quantized data, and the model iteratively applies denoising to recover the original image.
Because the diffusion model architecture does not need to be modified to accept new conditioning signals, diffusion dequantization methods generally allow greater freedom in model choice, for instance, by leveraging pretrained diffusion backbones~\cite{relic2024Lossya}.
They are also inherently multi-rate, as quantization strength is tied to the number of diffusion steps, which can be flexibly chosen at inference time.

Most related to our work, MRDIC~\cite{xu2025Decouplea} explores spatial adaptivity in image diffusion codecs.
They selectively prune image hyperlatents, removing features in non-salient areas, and densify and integrate these features to a conditional diffusion model using a transformer module and ControlNet~\cite{zhang2023Addingb}, respectively.
Due to the additional heavyweight modules, their method is significantly more computationally expensive than other diffusion image codecs.

Unlike previous work, we propose the first spatially adaptive codec within the diffusion dequantization paradigm~\cite{relic2025Bridging}.
By extending the diffusion model and entropy coding process to be ROI-aware in this framework, we avoid the need for heavyweight auxiliary modules and thus benefit from a smaller model size and inherent multi-rate functionality.

\section{Background}
\label{sec:background}
\subsection{Gaussian Diffusion}
Diffusion models~\cite{sohl-dickstein2015Deepa, ho2020Denoisingb, kingma2021Variational} define a process that progressively transitions between a data sample \(\mathbf{y}_0\) and pure noise \(\mathbf{y}_T\) over a series of \(T\) steps.
The forward process~(data-to-noise) can be expressed as a conditional distribution \(q(\mathbf{y}_t|\mathbf{y}_{t-1})\), which adds a small amount of Gaussian noise at each iteration.
Since this process is linear and Gaussian, the transition from the clean signal to any timestep~\(q(\mathbf{y}_t|\mathbf{y}_0)\) admits a closed-form solution:
\begin{equation}
\label{eq:gaussian_forward}
    \mathbf{y}_t= \alpha_t \mathbf{y}_0 +\sigma_t\epsilon, \quad \epsilon \sim \mathcal{N}(0, \mathbf{I}). 
\end{equation}
\(\alpha_t \text{ and } \sigma_t\) are strictly positive scalars representing the noise schedule.
This noise schedule, defined as a function of \(t\) (\ie \(\alpha_t = \alpha(t)\)), dictates the signal-to-noise~(SNR) ratio of \(\mathbf{y}_{t}\) at every timestep.

The reverse process~(noise-to-data) is intractable.
However, it can be estimated using a parameterized model \(p_{\theta}\) that learns \(p_{\theta}(\mathbf{y}_{t-1}|\mathbf{y}_t)\).
In practice, this model is commonly implemented as a denoising neural network \(p_{\theta}(\mathbf{y}_t, t)\) conditioned on the timestep \(t\).
Therefore, given some noisy data \(\mathbf{y}_t\), the clean signal \(\mathbf{y}_0\) can be recovered by iteratively applying the denoising model for \(t\) steps:
\begin{equation}
\label{eq:sampling}
    \mathbf{y}_{i-1} = p_{\theta}(\mathbf{y}_i, i), \quad i \in \{t,t-1 ...,1\}.
\end{equation}

\subsection{Uniform Diffusion}
Rather than Gaussian noise, some works have investigated diffusion models under non-Gaussian distributions~\cite{nachmani2021Denoising, heitz2023Iterativea, yang2024Progressive}.
Yang~\etal~\cite{yang2024Progressive} introduced universally quantized diffusion models, where the transition between timesteps is uniformly distributed, and formally derived the conditional distributions between transition states (\ie, \(q(\mathbf{y}_{t-1}|\mathbf{y}_t, \mathbf{y}_0)\)).
However, their formulation does not yield a simple distributional form for \(q(\mathbf{y}_t|\mathbf{y}_0)\) as exists in the Gaussian case.

Relic~\etal~\cite{relic2025Bridging} showed that, in practice, \(q(\mathbf{y}_t|\mathbf{y}_0)\) can be realized by replacing the standard normal noise in \cref{eq:gaussian_forward} with zero mean, unit variance uniform noise:
\begin{equation}
\label{eq:uniform_forward}
    \mathbf{y}_t = \alpha_t \mathbf{y}_0 + \sigma_t\mathbf{u}, \quad \mathbf{u} \sim \mathcal{U}\left(-\sqrt{3}, \sqrt{3}\right)^d,
\end{equation}
where \(d\) is the dimensionality of the data.
Denoising is performed following \cref{eq:sampling}.
They further demonstrated that such uniform noise diffusion models can be efficiently obtained by finetuning a Gaussian diffusion model on this uniform distribution~\cite{relic2025Bridging}, and build an image codec utilizing this model.

\vspace{-1em}
\subsection{Spatially Varying Diffusion}
Although not commonly used in practice, diffusion models can also be applied when the variance of added noise is independent per pixel~\cite{pearl2023SVNR, sahoo2024Diffusion}.
Sahoo~\etal~\cite{sahoo2024Diffusion} formalize a multivariate Gaussian noise schedule, where the scalar values \(\alpha_t \text{ and } \sigma_t\) are replaced by their vector counterparts. This yields the forward process:
\def\balpha{\boldsymbol{\alpha}}
\def\bsigma{\boldsymbol{\sigma}}
\begin{equation}
    \mathbf{y}_t= \balpha_t \mathbf{y}_0 +\text{diag}(\bsigma_t)\epsilon, \quad \epsilon \sim \mathcal{N}(0, \mathbf{I}),
\end{equation}
where \(\mathbf{x}_t, \mathbf{x}_0 \in \mathbb{R}^d,\text{ and }\balpha_t, \bsigma_t \in \mathbb{R}_+^d\)~\cite{sahoo2024Diffusion}.
Compared to standard Gaussian diffusion, \(\balpha_t \text{ and }\bsigma_t\) dictate the SNR ratio of \emph{each pixel} as a function of \(t\), rather than the entire image.

\section{Method}
\label{sec:method}
\def\bt{\mathbf{t}}

\begin{figure*}[t!]
    \centering
    \includegraphics[width=.95\linewidth]{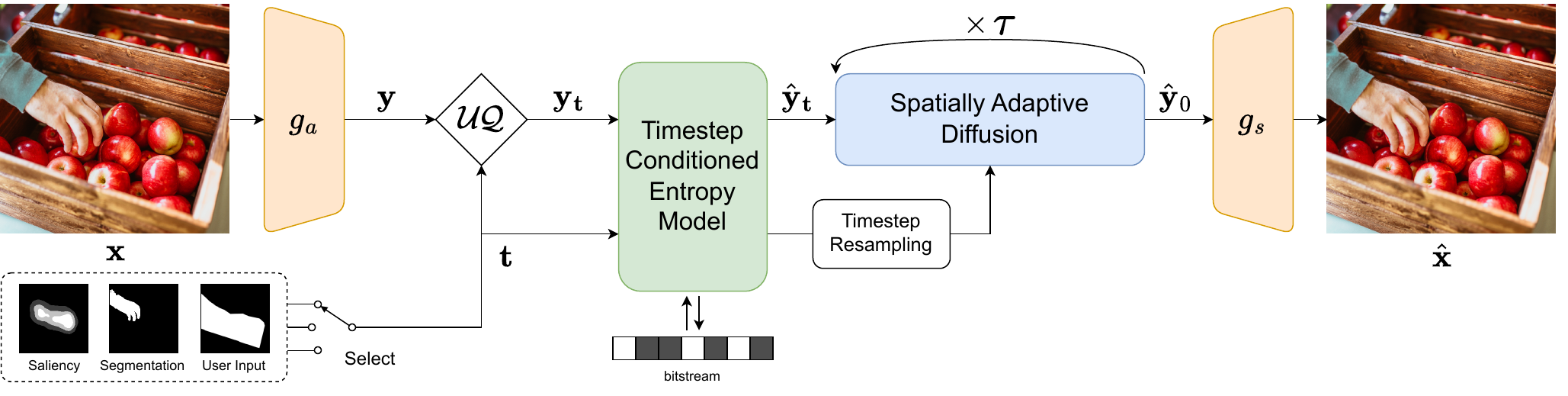}
    \caption{Pipeline of our proposed method. An image \(\mathbf{x}\) and ROI map \(\mathbf{t}\) are taken as input and the former encoded to latent space with a VAE encoder. The image latent \(\mathbf{y}_{\mathbf{t}}\) is spatially adaptively quantized according to the ROI map, adding noise in the process. \(\mathbf{y}_{\mathbf{t}}\) and \(\mathbf{t}\) are then transmitted to the bitstream using our proposed timestep conditioned entropy model. At the receiver, \(\mathbf{y}_{\mathbf{t}}\) and \(\mathbf{t}\) are processed with our spatially adaptive diffusion model, including timestep resampling, which reconstructs the source latent. It is then decoded with a VAE decoder to produce the final reconstruction \(\hat{\mathbf{x}}\).}
    \label{fig:pipeline}
\end{figure*}

We first introduce a spatially varying uniform diffusion process~(\cref{subsec:spatial_diff}), then apply it to compression by designing a region-adaptive generative image codec based on this process~(\cref{subsec:compression_pipeline}).

 \vspace{-0.5em}
\subsection{Spatially Varying Uniform Diffusion}
\label{subsec:spatial_diff}
We define a discrete-time diffusion process with a multivariate uniform noise schedule.
Our goal is to use this process to introduce noise via quantization~\cite{relic2025Bridging}; we are therefore concerned with defining the forward process \(q(\mathbf{y}_{\bt}|\mathbf{y}_0)\), expressed functionally as:
\begin{equation}
\label{eq:spatial_uniform_forward}
    \mathbf{y}_{\bt} = \balpha_{\bt}\mathbf{y}_0 + \bsigma_{\bt}\mathbf{u}, \quad \mathbf{u} \sim \mathcal{U}\left(-\sqrt{3}, \sqrt{3}\right)^d.
\end{equation}
Here, \(\balpha_{\bt} \text{ and }\bsigma_{\bt}\) are indexed with a \emph{spatial timestep map} \(\bt \in \mathbb{R}_+^d\) and defined for every pixel in \(\mathbf{y}_{\bt}\).
Intuitively, each pixel can be at its own timestep throughout the diffusion process.

To support entropy coding and practical use in an image codec, we follow the diffusion dequantization paradigm~\cite{relic2025Bridging} and use universal quantization~\cite{zamir1992universal} in the forward diffusion process, rather than a linear combination with sampled noise:
\begin{equation}
\label{eq:our_forward}
\begin{gathered}
    \hat{\mathbf{y}}_{\bt} = \left\lfloor \balpha_{\bt} \mathbf{y}_{0} - \mathbf{u} \right\rceil_{\Delta_{\bt}} + \mathbf{u}, \\
    \text{with }\mathbf{u} \sim \mathcal{U}\left(-\frac{\Delta_{\mathbf{t}}}{2}, \frac{\Delta_{\bt}}{2}\right),~\Delta_{\bt} = \sqrt{12 \bsigma_{\bt}^2},
\end{gathered}
\end{equation}
where \(\lfloor \cdot \rceil_{\Delta_{\mathbf{t}}}\) denotes quantization to a bin width of \(\Delta_{\mathbf{t}}\)~\cite{relic2025Bridging}.
In addition to representing the number of diffusion steps, \(\bt\) acts as a rate control hyperparameter; it dictates the quantization bin width, which determines the amount of information in each pixel in \(\hat{\mathbf{y}}_\bt\).

The reverse process is approximated via a denoising model \(p_{\theta}(\hat{\mathbf{y}}_{\bt}, \bt)\) conditioned on a noisy sample \(\hat{\mathbf{y}}_\bt\) and a per-pixel timestep map \(\bt\), rather than a uniform scalar timestep as in traditional diffusion formulations.

\subsubsection{Sampling Strategy.}
\label{subsec:decoder_sampling_strategy}
As defined in \cref{eq:our_forward}, each pixel in \(\hat{\mathbf{y}}_{\bt}\) is noised to some arbitrary timestep \(t\).
This implies that each pixel requires a different number of forward passes through the diffusion model to remove all noise.
Naively applying this scheme can introduce hard boundaries between regions with significantly different noise levels, thereby breaking properties of the underlying noise distribution~\cite{chang2023How}.

\begin{figure}[t]
    \centering
    \begin{subfigure}{0.54\linewidth}
        \includegraphics[width=0.99\linewidth]{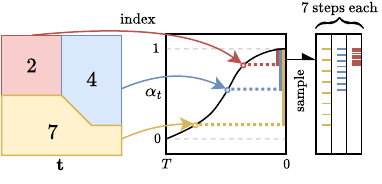}
        \caption{}
        \label{fig:sampling}
    \end{subfigure}
    \begin{subfigure}{0.45\linewidth}
        \includegraphics[width=0.99\linewidth]{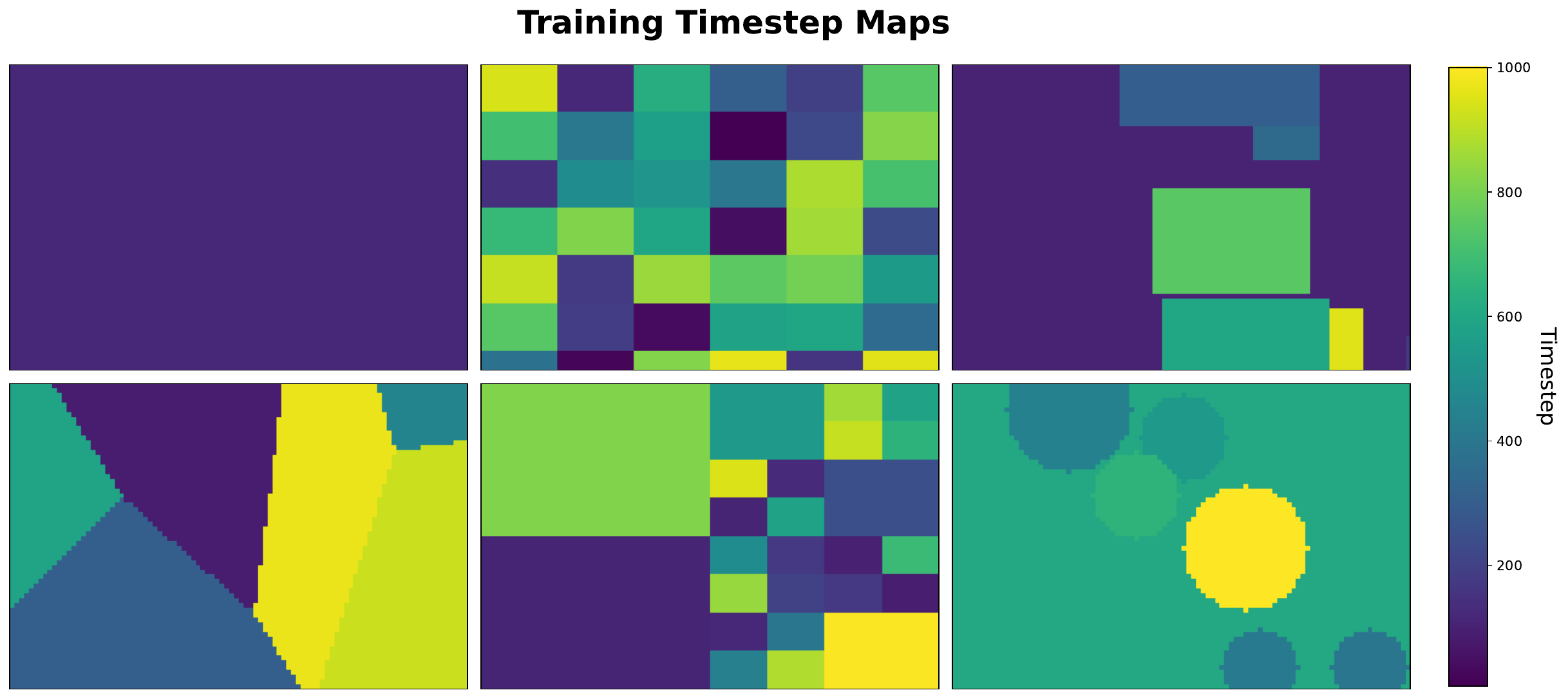}
        \caption{}
        \label{fig:train_t}
    \end{subfigure}
    \caption{\textbf{(a)} Our timestep resampling process. For each unique noise level of pixels in \(\hat{\mathbf{y}}_{\bt}\)~(denoted by different colors), we adjust the am noise removed per step such that all pixels are fully denoised in the same number of diffusion forward evaluations. \textbf{(b)} Examples of the spatial \(\bt\) maps we sample during training. The region shape, number of regions, and timestep value of each region is randomly sampled.}
\end{figure}

To solve this issue, we denoise the entire sample \(\hat{\mathbf{y}}_{\bt}\) in the same number of diffusion iterations, despite the varying timesteps of the pixels within.
In practice, we identify that the noise level defined by \(\bt\) is independent of how many denoising steps it takes to a pixel to be fully denoised; the latter only determines the step size of each denoising iteration.
Thus, we scale the denoising rate of each pixel such that the entire image is denoised with the same number of diffusion forward passes, which we call timestep resampling~(shown in \cref{fig:sampling}).
We first identify a number of diffusion iterations \(\tau = \text{max}(\bt)\) to perform.
Then, using each pixel's initial noise level, we resample the domain of \(\balpha_{\bt} \text{ and } \bsigma_{\bt}\) such that every pixel takes \(\tau\) steps to reach the fully denoised state.
This process gradually denoises all pixels together, reducing boundary artifacts between regions with different noise levels.

\subsection{Compression pipeline}
\label{subsec:compression_pipeline}
\cref{fig:pipeline} shows our proposed compression pipeline.
It takes an input image \(\mathbf{x}\) and a timestep map \(\bt\), denoting which areas of the input image to preserve and which can tolerate larger distortions.
\(\bt\) can be a user input, or computed from \(\mathbf{x}\) via saliency prediction or segmentation model, for example.

At the encoder, \(\mathbf{x}\) is first encoded to the latent space via a VAE encoder \(g_a\).
It is then quantized to spatially varying uniform noise levels using~\cref{eq:our_forward}.
This quantized latent \(\hat{\mathbf{y}}_t\) is then encoded into a bitstream via our proposed timestep conditioned entropy model~(see below).

At the receiver, \(\hat{\mathbf{y}}_t\) is denoised by our spatially varying uniform diffusion model to produce a clean latent \(\hat{\mathbf{y}}_0\), following the sampling strategy discussed in \cref{subsec:decoder_sampling_strategy}.
The clean latent is decoded to image space by a VAE decoder \(g_s\) to obtain the final reconstructed image \(\hat{\mathbf{x}}\).

\subsubsection{Timestep Conditioned Entropy Model}
\label{subsec:timestep_conditioned_entropy_model}
To perform the correct number of denoising steps during decoding, our codec requires \(\bt\) to be transmitted to the receiver~\cite{relic2025Bridging}, which incurs additional rate overhead.
However, we identify that \(\bt\) dictates the bin width used during quantization, and therefore encodes prior information on the distribution of \(\hat{\mathbf{y}}_{\bt}\).
Previous work shows that prior knowledge of the latent distribution can improve compression performance~\cite{balle2018Variationala, minnen2018Jointa, he2022ELICa}.
Therefore, we propose to use \(\bt\) as explicit context for entropy coding and introduce a timestep conditioned entropy model.

\begin{figure}[t!]
    \centering
    \includegraphics[width=0.9\textwidth]{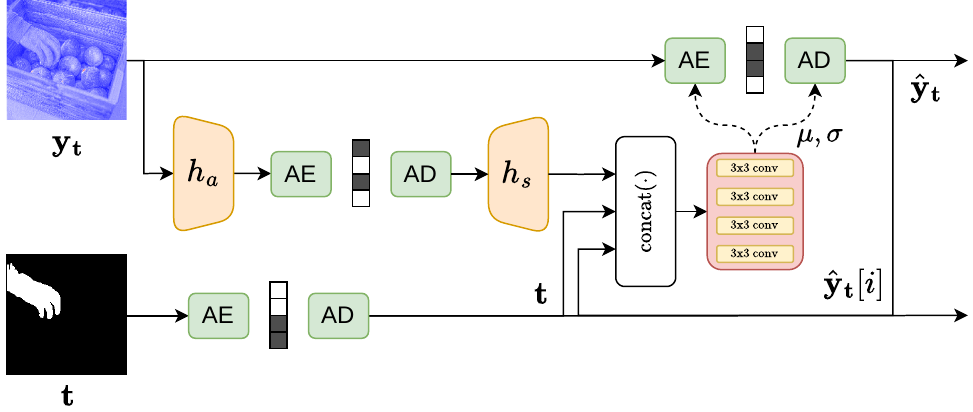}
    \caption{Our Timestep Conditioned Entropy Model architecture. The ROI map \(\mathbf{t}\) is first losslessly encoded and transmitted to the receiver. It is then used as prior knowledge for entropy coding, along with hierarchical and channel-slice context, by predicting Gaussian parameters that define the distribution of \(\mathbf{y}_{\mathbf{t}}\).}
    \label{fig:em}
\end{figure}

We implement this using a dual-stream architecture, shown in \cref{fig:em}.
In the first stream, \(\bt\) is losslessly compressed and transmitted to the receiver.
The second stream compresses the latent \(\mathbf{y}_{\bt}\) conditioned on \(\bt\).
We combine \(\bt\) with additional hierarchical~\cite{balle2018Variationala} and channel-slice~\cite{minnen2018Jointa} priors via concatenation.
This joint context is then passed to a convolutional encoder that predicts means and scales of a Gaussian that parameterizes the latent distribution, which is used in entropy coding.

\section{Implementation details}

\subsubsection{Architecture}
\label{subsec:architecture}
Following previous works~\cite{ho2020Denoisingb, rombach2022HighResolutiona}, we implement our diffusion denoising model as a time-conditional U-Net~\cite{ronneberger2015UNet}.
Training this architecture from scratch, while possible, is computationally prohibitive without large-scale resources and budget~\cite{rombach2022HighResolutiona, kingma2021Variational, ho2020Denoisingb}.
We thus initialize our spatially varying diffusion model with the weights of a spatially uniform foundation model~(\ie, Stable Diffusion~\cite{rombach2022HighResolutiona}), thereby significantly reducing required training compute.

The main difference between the two architectures is the use of a spatial \(\bt\) or a scalar \(t\); we must embed timestep information per pixel rather than across the whole image.
Standard U-Net-based diffusion models compute timestep embeddings by passing \(t\) to a fully connected embedding module and broadcasting them with the intermediate U-Net features.
We replace all linear timestep embedding layers with pixelwise convolutions, effectively computing timestep embeddings per pixel.
When necessary, we average pool the embeddings to match the spatial shape of the U-Net features.
Because these operations are mathematically equivalent across the channel dimension, we directly initialize the convolutional filters with the pretrained linear layer weights.

\subsubsection{Training}
We train the diffusion network and entropy model of our method in two separate stages.
Our diffusion model is fully initialized using weights from Stable Diffusion v2.1~\cite{rombach2022HighResolutiona}.
We then finetune it for our task by replacing Gaussian noise with uniform noise~\cite{relic2025Bridging} and using spatial timestep maps.
We generate such \(\bt\) maps by randomly placing rectangular, circular, grid, or voronoi regions of varying \(t\) values on top of a constant \(t\) canvas~(\cref{fig:train_t}).
Constant~(\ie, spatially uniform) maps are chosen with a probability of \(0.15\).
Our model is optimized for 200k steps with batch size 32 and learning rate \(10^{-5}\) using the standard \(\mathcal{L}_{\text{simple}}\) diffusion loss~\cite{ho2020Denoisingb} with \(\mathbf{v}\)-prediction~\cite{kingma2021Variational}.
We then freeze the diffusion model and train the entropy model with a rate objective \(\mathcal{L}_{\text{rate}}=-\log_2p_{\hat{\mathbf{y}}_{\bt}}(\hat{\mathbf{y}}_{\bt})-\log_2p_{\hat{\mathbf{z}}}(\hat{\mathbf{z}})\), where \(\hat{\mathbf{z}}\) is the hyperlatent~\cite{balle2018Variationala} and \(p_{\hat{\mathbf{y}}_{\bt}}\text{ and } p_{\hat{\mathbf{z}}}\) are predicted probability distributions over the latents and hyperlatents, respectively.
The entropy model is trained on image crops ($512\times512$) from the LSDIR dataset~\cite{li2023LSDIR} for 375k steps with a batch size of 8 and a learning rate of \(10^{-4}\).

\section{Experiments}
\label{sec:experiments}

Our method bridges two distinct paradigms: spatially adaptive coding and generative compression.
Existing approaches in these fields prioritize different goals; the former focus on pixelwise distortion metrics across spatial regions, while generative codecs emphasize global perceptual quality.
As no available baseline addresses both domains simultaneously\footnote{The closest related work MRIDC~\cite{xu2025Decouplea} does not have publicly available code or reconstructions to compare against.}, we benchmark our proposed method against both domains independently.

\subsubsection{Spatially Adaptive Methods.}
Spatially adaptive methods remain the most direct comparison with our proposal.
We compare our method the current state-of-the-art controllable ROI baselines~\textbf{C-ROI}~\cite{jin2025Customizable} and Kao~\etal~\cite{kao2023TransformerBased}~(referred to as \textbf{TV-ROI}).
Notably, these methods are not generative and prioritize pixelwise distortion metrics such as PSNR, which are known to be ineffective measures of perceptual quality~\cite{blau2019Rethinking, mentzer2020HighFidelitya}.
Therefore, we also build and compare against a spatially adaptive generative variation of UDDQ~(referred to as \textbf{UDDQ-ROI}).
We take the spatially uniform diffusion codec presented in Relic~\etal~\cite{relic2025Bridging} and apply their proposed quantization scheme per-pixel, according to the ROI map, enabling spatial bitrate control.
The rest of the pipeline~(\ie, entropy and diffusion models) remains unchanged, and we set the number of diffusion steps to achieve the best quality in the in-ROI.~\ranote{We could also add visual results on the supplementary to the case in which we set the number of steps to better reconstruct the non-ROI. Probably enough to leave it out of the main paper.}
To quantitatively evaluate these methods, we use \textbf{LPIPS} computed on in-ROI regions and the entire image, and plot rate-distortion performance.

\begin{figure}[t]
    \centering
    \includegraphics[width=0.99\linewidth]{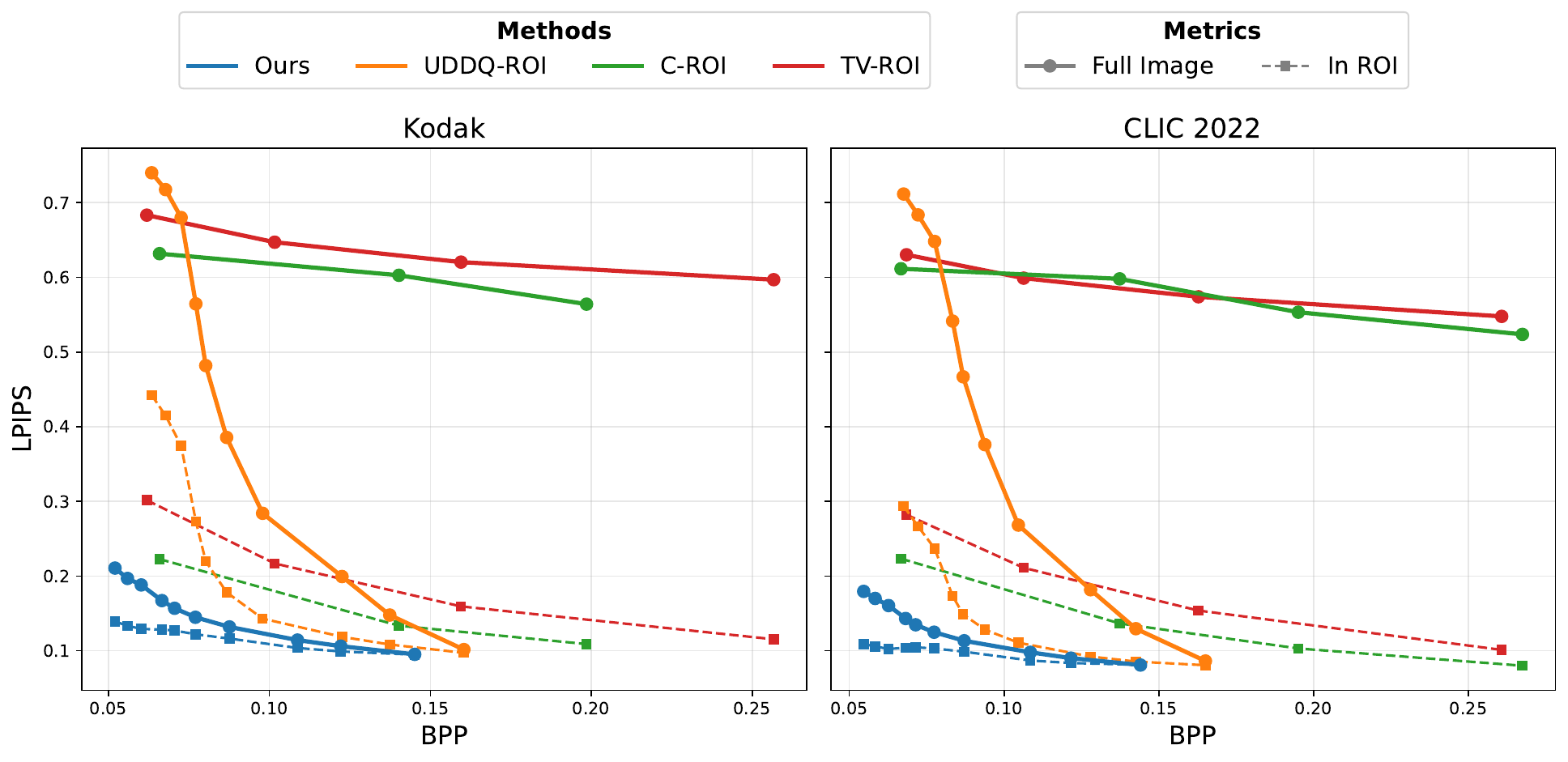}
    \caption{Quantitative rate-distortion comparison of spatially adaptive baselines on the Kodak and CLIC 2022 datasets. Performance is measured by full image (solid line) and in-ROI (dotted line) LPIPS, where our method is best performing in both metrics. UDDQ-ROI is competitive at high bitrates, but quickly degrades as bitrate decreases, while the other baselines do not show strong performance in perceptual metrics.}
    \label{fig:roi_rd}
\end{figure}

\subsubsection{Generative Methods.}
For a fair evaluation with other perceptually-oriented codecs, we also compare with generative, spatially uniform methods.
We choose Stable Diffusion-based generative codecs as baselines, specifically \textbf{DiffEIC}~\cite{li2025Extreme}, \textbf{PerCo (SD)}~\cite{korber2024PerCo}~(an open source implementation of PerCo~\cite{careil2024image}), and Relic~\etal~\cite{relic2025Bridging}~(further referred to as \textbf{UDDQ}).
We also use \textbf{ILLM}~\cite{muckley2023Improving} as the SoTA GAN-based generative method, and use \textbf{VTM}~\cite{jointvideoexpertsteamjvet2020VVC} as a reference to the latest standardized traditional image codec.
As the main motivation of our work is to adapt generative compression to reflect human visual attention, we conduct a user study to evaluate whether our spatially adaptive codec is preferred to existing spatially uniform methods.
Each of the above baselines is compared pairwise with our method.
Participants are shown two images, as well as the ground truth, and asked to indicate their strong or weak preference for one of the images.
We selected images from the Kodak dataset with comparable bitrates in all methods, resulting in 13 samples between \(0.03\) and \(0.1\) bpp.
In total, we collected 745 ratings from 16 participants.
Additional details of our user study are listed in the supplemental material, as well as quantitative rate-distortion analysis.

\subsubsection{Datasets.}
We use the \textbf{Kodak}~\cite{franzenKodak} and \textbf{CLIC 2022}~\cite{toderici2022CLIC} test datasets for all comparisons.
As ground truth masks are not available for these datasets, we generate ROI maps for both datasets by manually segmenting foreground, salient, or otherwise visually important regions in each image.
We choose to create ROI masks for compression datasets, rather than use other datasets with ground truth segmentation masks~(\eg, COCO), as these datasets use a lossy storage format which introduces artifacts that are not present in natural images and can skew evaluation metrics.
We emphasize that our method can take any arbitrary ROI mask as input; we predefine them here for fair evaluation.

\section{Results}
\label{subsec:results}

\subsubsection{Spatially Adaptive Methods.}

\begin{figure}[t]
    \centering
    \begin{subfigure}{\linewidth}
    \centering
        \includegraphics[width=0.9\linewidth]{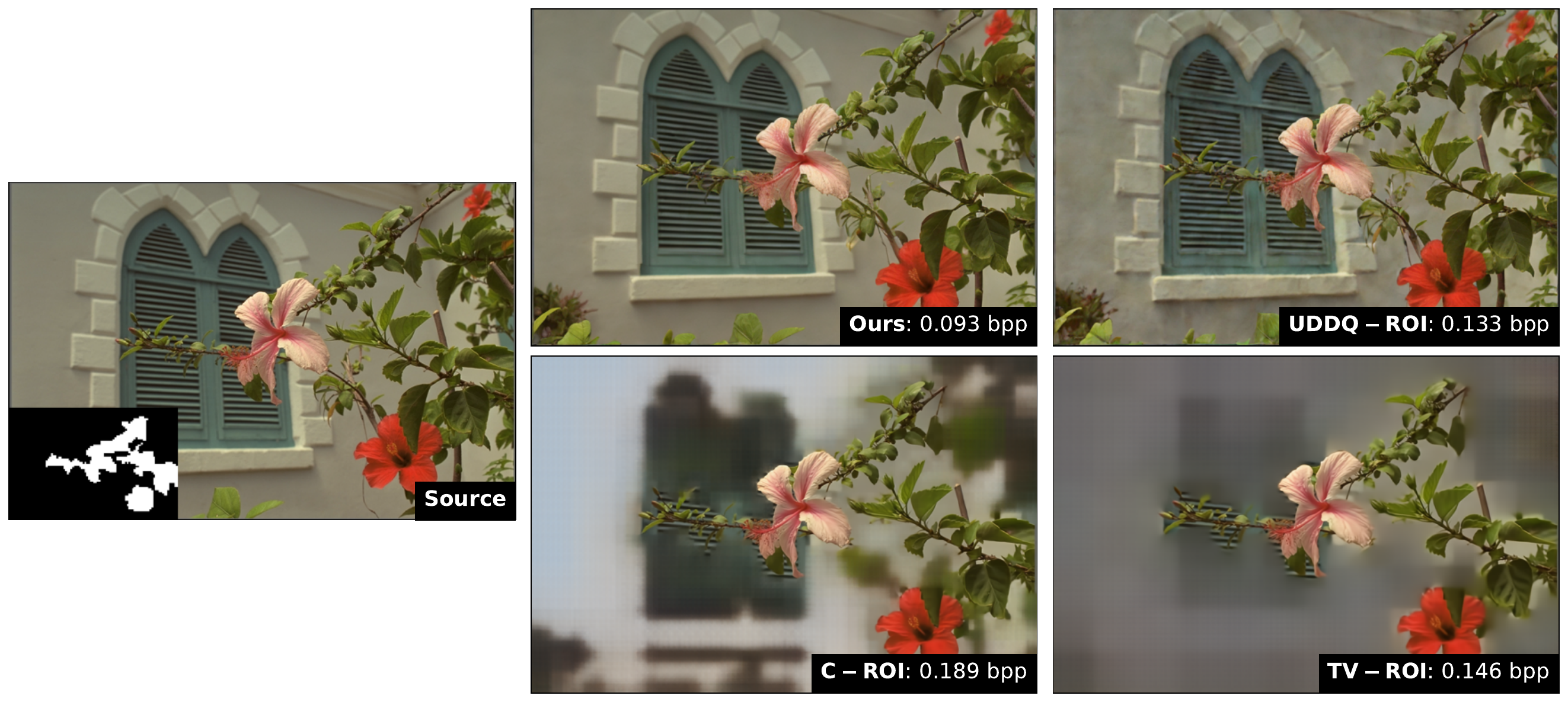}
    \end{subfigure}
    \begin{subfigure}{\linewidth}
    \centering
        \includegraphics[width=0.9\linewidth]{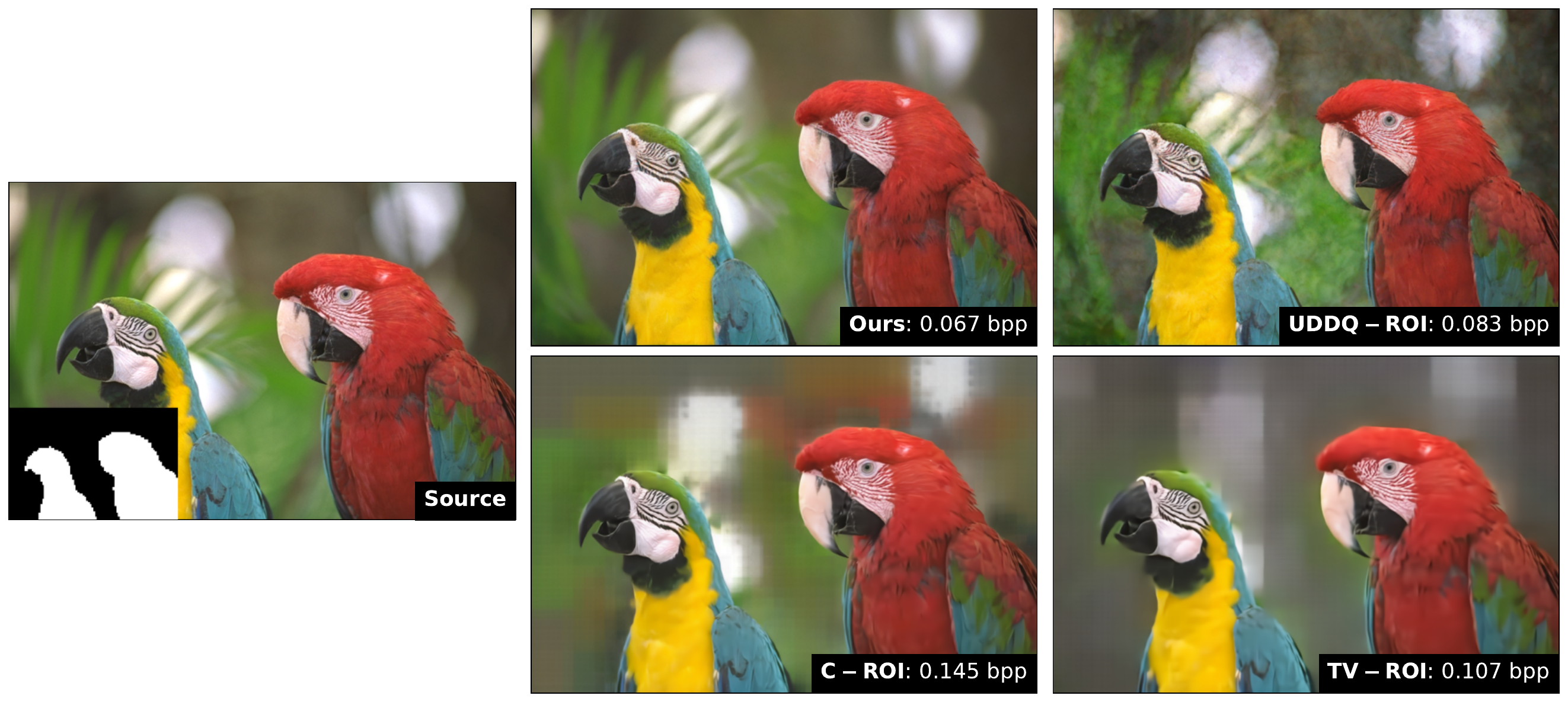}
    \end{subfigure}
    \caption{Visual examples of spatially adaptive methods (C-ROI~\cite{jin2025Customizable} and TV-ROI~\cite{kao2023TransformerBased}) on the Kodak dataset. All methods accurately reconstruct ROI content. In non-ROI areas, UDDQ-ROI displays unnatural artifacting and artificial textures, and the other methods have a significant lack of detail. Additionally, our method has no boundary artifacts betweeen ROI and non-ROI areas.}
    \label{fig:roi_qualitative}
\end{figure}

Quantitative rate-distortion comparisons are shown in \cref{fig:roi_rd}.
The non-generative baselines, C-ROI and TV-ROI, allocate a large majority of their bits to the ROI.
Although this achieves reasonable local quality, it results in a substantial difference between their ROI-masked and global LPIPS scores.
Conversely, UDDQ-ROI prioritizes perceptual quality, making its LPIPS scores more competitive within the ROI at higher bitrates; however, its performance significantly degrades as bitrate decreases.
Our method addresses these limitations and outperforms all baselines in both full-image and ROI-masked LPIPS. 
It maintains high perceptual quality throughout the image and remains consistent in the low bitrate regime.

\cref{fig:roi_qualitative} shows visual examples of these methods.
As reflected in the quantitative metrics, C-ROI and TV-ROI exhibit significant degradation in the background areas.
Because these methods lack a generative prior, they are unable to synthesize realistic textures from the heavily quantized representations outside the ROI, leading to severe blurring and loss of detail.
Furthermore, while UDDQ-ROI can reconstruct the background content, it introduces distinct, unnatural textural artifacts.
This occurs because the baseline architecture was not natively designed for spatial adaptivity, highlighting the need for specific architectural considerations when building spatially adaptive generative codecs.
In contrast, our method seamlessly integrates regions inside and outside the ROI without introducing boundary artifacts.
The reconstructions are detailed within the ROI, while the generative prior ensures that the background regions remain realistic and semantically plausible.

\begin{figure}[t]
    \centering
    \begin{subfigure}{0.55\linewidth}
        \includegraphics[width=0.95\linewidth]{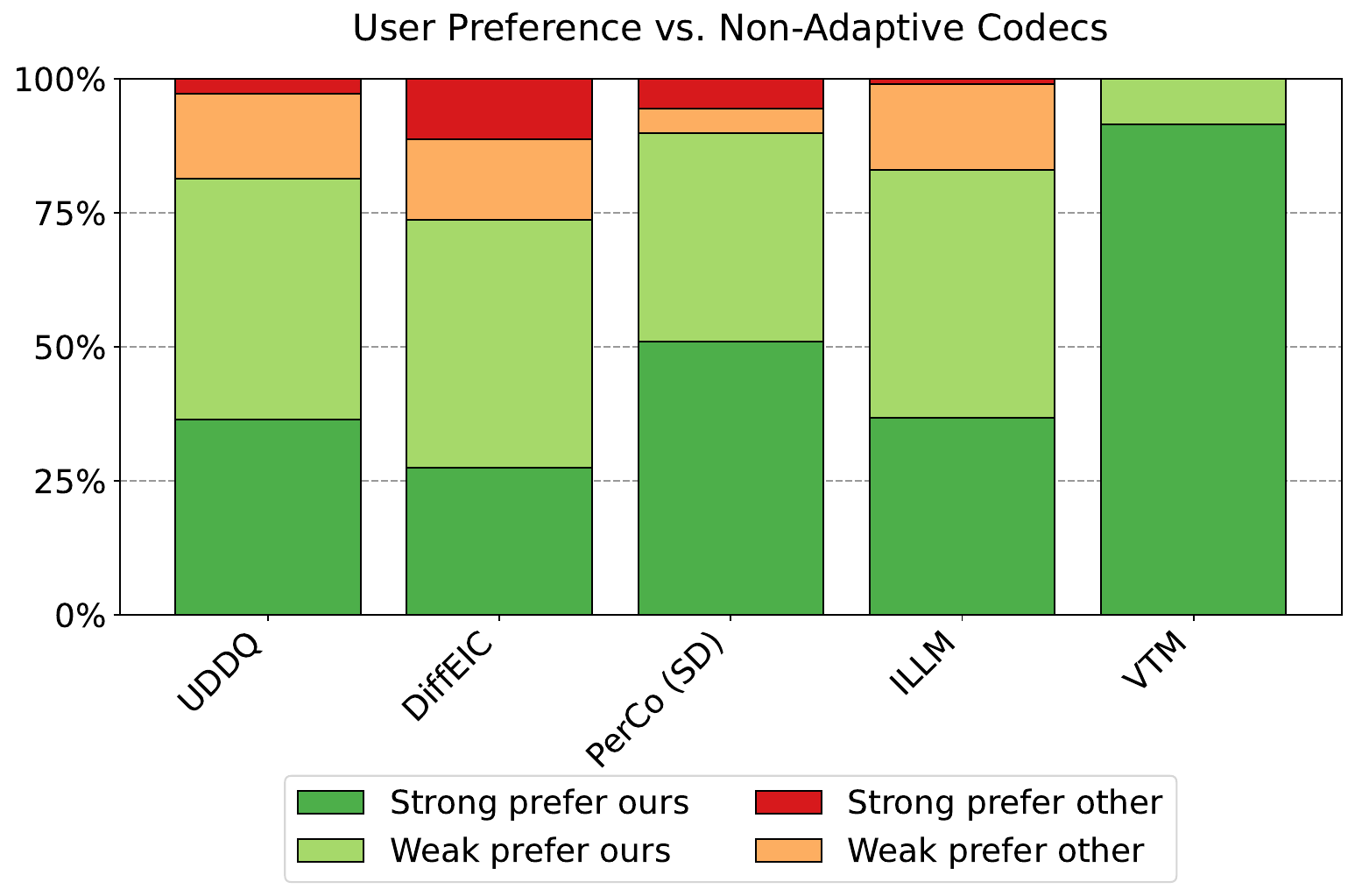}
        \caption{}
        \label{fig:user_study}
    \end{subfigure}
    \begin{subfigure}{0.425\linewidth}
        \includegraphics[width=0.99\linewidth]{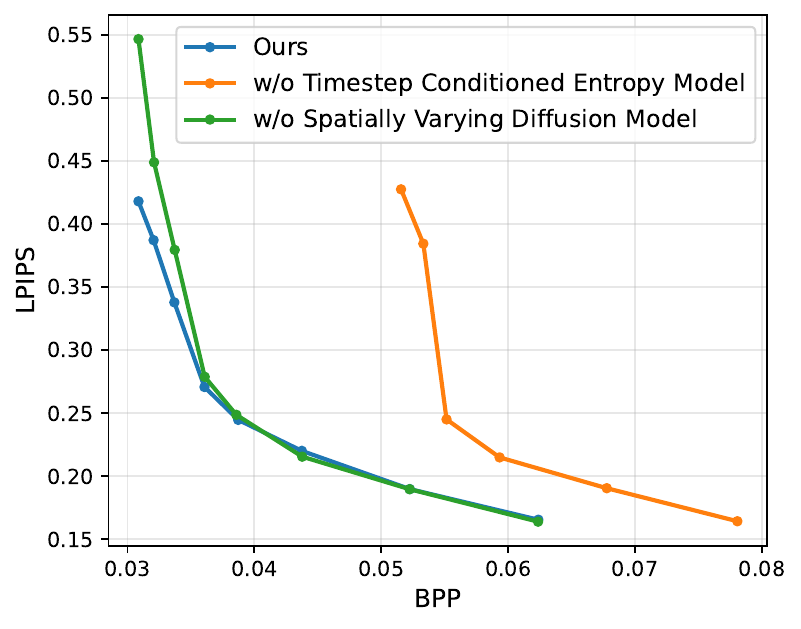}
        \caption{}
        \label{fig:ablation_rd}
    \end{subfigure}
    \caption{\textbf{(a)}~User study results of our method with non-spatially adaptive baselines. Our method is generally preferred against all other baselines, as well as in \(85\%\) of all comparisons. \textbf{(b)}~Rate-distortion performance comparisons of our ablation study. Our timestep conditioned entropy model is crucial to our method's performance, and our spatially varying diffusion model improves image quality at low rates.}
\end{figure}

\subsubsection{Generative Methods.}
When considering the end user experience, our spatially adaptive reconstructions were preferred over other spatially uniform methods, as shown in \cref{fig:user_study}.
Across all baselines, our method was preferred in \(85\%\) of the comparisons and was strongly preferred in \(48\%\).
Generally, diffusion-based methods performed stronger than the GAN-based or traditional baseline.

Visual examples of reconstructions are shown in~\cref{fig:visual_comparison}.
Compared to the baselines, objects within the ROI~(e.g., the hand, house, and lighthouse) are reconstructed with greater detail and more faithful to the source image.
Areas outside an ROI, such as the rocks and statue shoulder, although potentially semantically different, appear consistently realistic and free from unnatural artifacts.

\def\picwidth{0.9\textwidth}
\begin{figure*}[t]
    \begin{subfigure}{\textwidth}
        \centering
        \includegraphics[width=\picwidth]{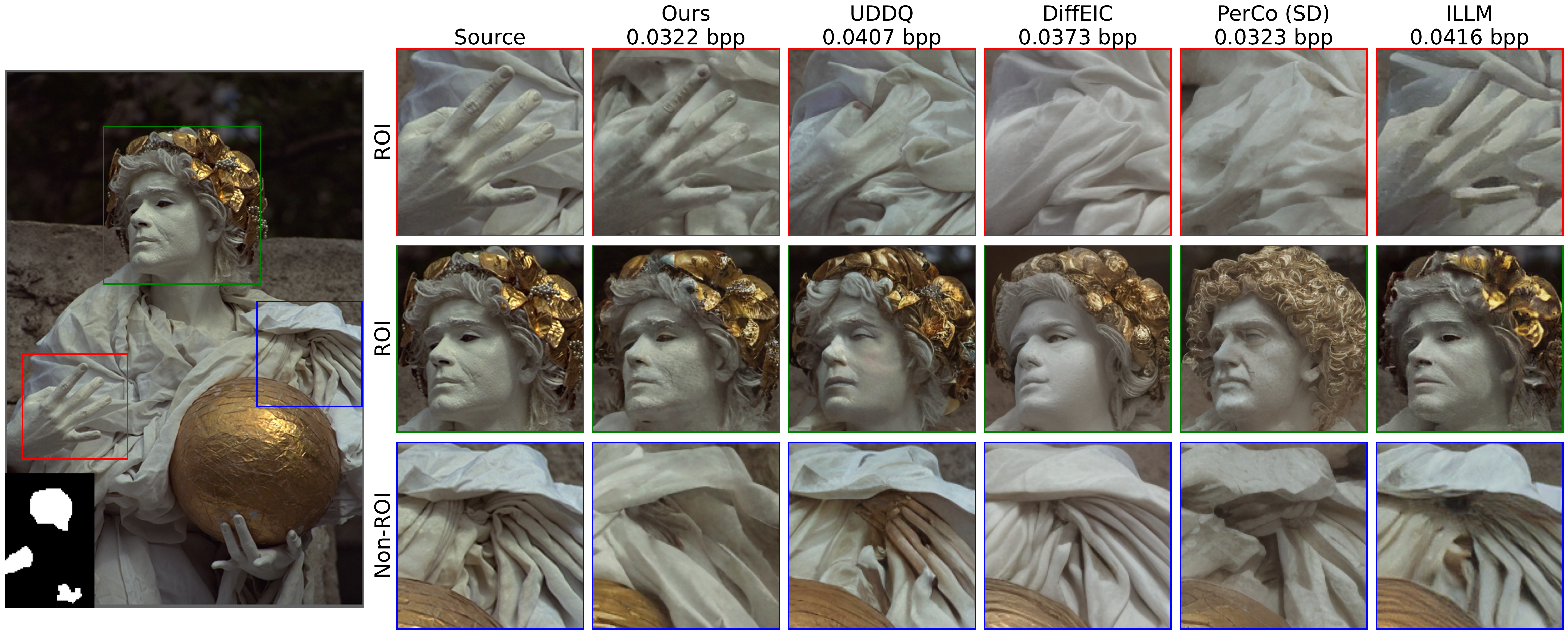}
    \end{subfigure}
    \vspace{2em}
    \begin{subfigure}{\textwidth}
        \centering
        \includegraphics[width=\picwidth]{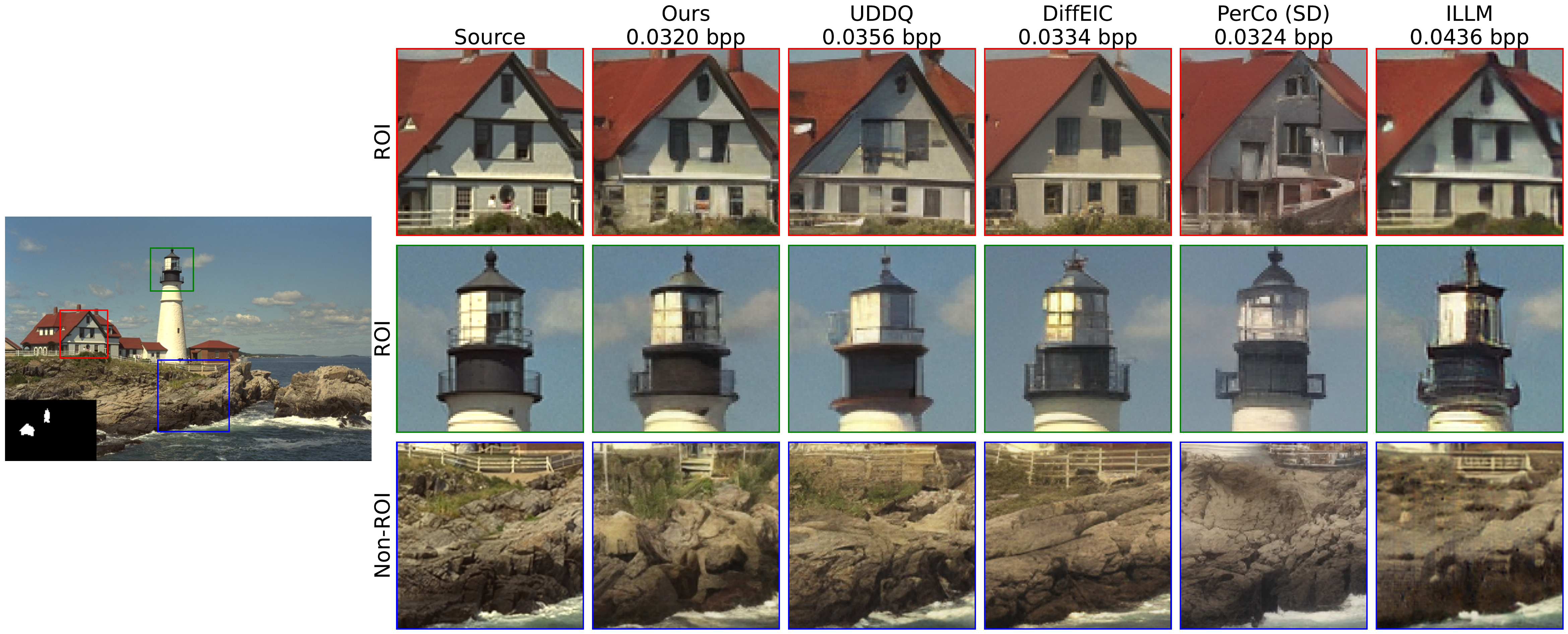}
    \end{subfigure}
    \caption{Qualitative comparison of our method to non-spatitally adaptive generative baselines (UDDQ~\cite{relic2025Bridging}, DiffEIC~\cite{li2025Extreme}, PerCo~\cite{careil2024image}, ILLM~\cite{muckley2023Improving}). Our method produces faithful reconstructions in ROI areas while maintaing a realistic appearance in non-ROI areas, even if the content may slightly differ. Best viewed digitally.}
\label{fig:visual_comparison}
\end{figure*}

\subsection{Additional Studies.}
\paragraph{Ablation studies.}
We validate the effectiveness of our proposed changes by performing the following ablation studies: \textit{1)} removing timestep conditioning from our entropy model, and \textit{2)} replacing our spatially varying diffusion model with RePaint~\cite{lugmayr2022RePainta}, an inpainting method using pretrained Stable Diffusion\footnote{Exact details can be found in the supplementary material.}~\cite{rombach2022HighResolutiona}.
Experiments are performed on the Kodak dataset and measured with LPIPS, shown in \cref{fig:ablation_rd}.
Removing our proposed timestep conditioned entropy model results in significantly higher bitrates to achieve the same perceptual quality, indicating it is crucial to the compression performance of our method.
We also observe increased training stability when using timestep conditioning.
Replacing spatial diffusion with RePaint decreases perceptual quality at lower bitrates, validating the use of our proposed diffusion process for low-rate compression.

\begin{figure}[t]
    \centering
    \includegraphics[width=0.95\linewidth]{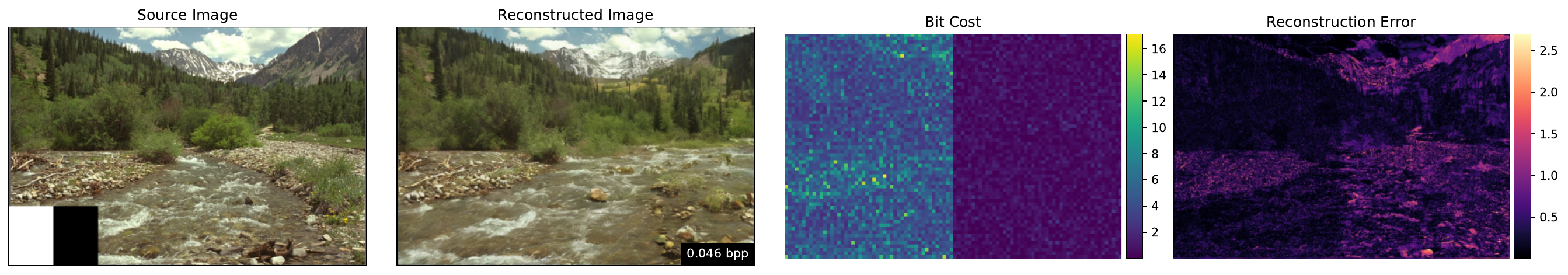}
    \caption{Spatial bit allocation of our method. More bits are spent transmitting areas in-ROI, resulting in lower reconstruction errors, while the inexpensive out-ROI areas maintain high realism and do not contain boundary artifacts.}
    \label{fig:rate_allocation}
\end{figure}

\paragraph{Parameter Efficiency.}
We evaluate the parameter efficiency of our method and the diffusion baselines by computing model sizes.
UDDQ, which follows a similar paradigm but for spatially uniform compression, is the most compact with 949.8M parameters.
Because we achieve spatial adaptivity natively, our method requires only 953.7M parameters~(the second smallest).
This marginal increase of roughly 4M parameters over UDDQ stems from our timestep-conditioned entropy model, as both methods use the same diffusion backbone.
In contrast, our closest spatially adaptive competitor, MRIDC, requires 1551.6M parameters due to its reliance on heavyweight feature sparsification and diffusion control modules.\footnote{As the implementation for MRIDC are not publicly available, our empirical comparison with this method is restricted to model size.}
Finally, baselines like DiffEIC and PerCo~(SD) require 1379.5M and 5041.1M parameters, respectively, stemming from auxiliary control networks and large text embedding models.

\paragraph{Bit Allocation.}
\cref{fig:rate_allocation} shows how our method dynamically allocates bits according to the input ROI mask.
More bits are spent transmitting in-ROI latents, which results in lower reconstruction error and a closer match to the source image.
As significantly fewer bits are allocated to non-ROI areas, the reconstruction error is higher.
However, the final image is still realistic and semantically plausible in these regions.

\section{Conclusion}
\label{sec:conclusion}

We present a region-adaptive generative image codec built around our proposed spatially varying diffusion model, enabling controllable bit allocation in generative compression.
Our method sets a new state-of-the-art in both ROI-masked and full image perceptual quality compared to other spatially adaptive codecs.
Future work includes: applying our codec to related contexts, such as foveated image compression, exploring region-adaptive diffusion in other image processing tasks, such as image editing or inpainting, and extending our spatially varying diffusion model into the temporal domain.

\pagebreak
\bibliographystyle{splncs04}
\bibliography{main}

\clearpage
\appendix
\begin{center}
{\Large\bfseries Region-Adaptive Generative Compression with Spatially Varying Diffusion Models}\\[1em]
{\large Supplementary Material}
\end{center}
\vspace{2em}
\def\bt{\mathbf{t}}
\def\balpha{\boldsymbol{\alpha}}
\def\bsigma{\boldsymbol{\sigma}}

\section{Additional Details}

\subsection{Visual Results}
\begin{figure*}[]
    \begin{subfigure}{\textwidth}
        \centering
        \includegraphics[width=\textwidth]{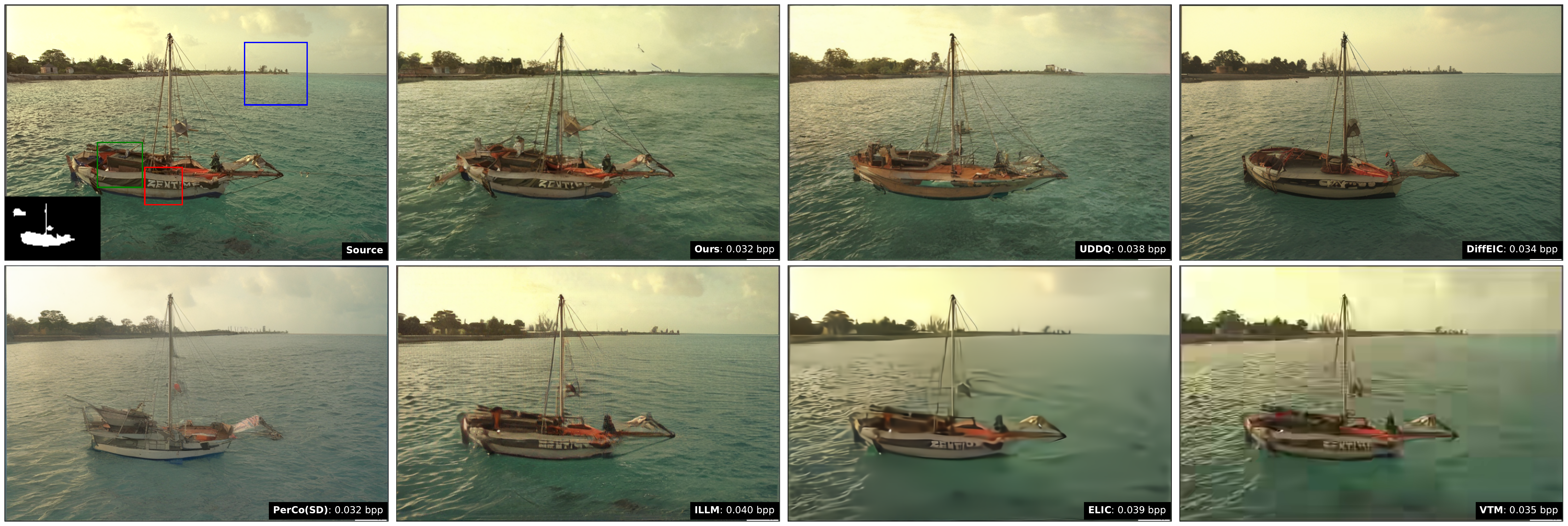}
    \end{subfigure}
    \begin{subfigure}{\textwidth}
        \centering
        \includegraphics[width=\textwidth]{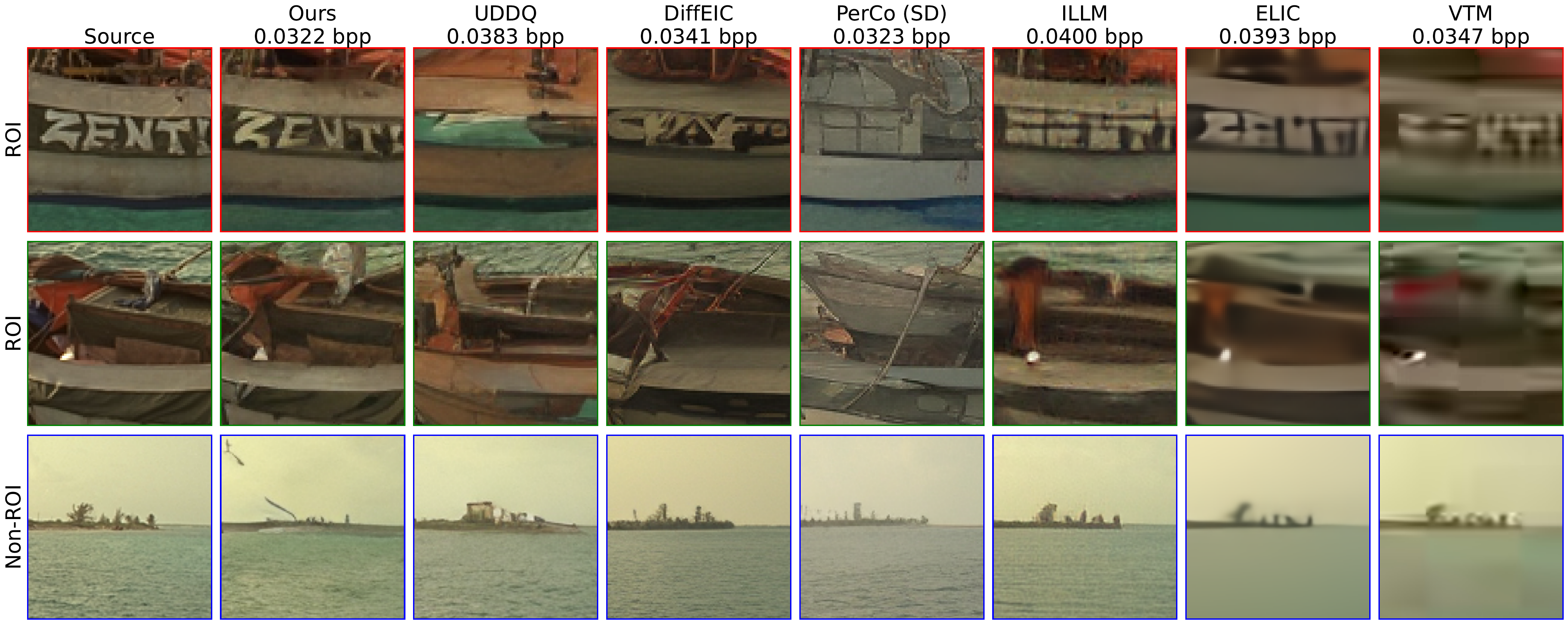}
    \end{subfigure}
    \caption{Full image (top) and zoomed-view (bottom) qualitative comparison between our method and the baselines. When considering the entire image, our method looks consistently realistic. In ROI areas, the reconstruction is faithful to the source, and areas outside of ROI can vary semantically but still remain plausible. Best viewed digitally.}
\label{fig:kodim06_full_and_crops}
\end{figure*}

\begin{figure*}[]
    \begin{subfigure}{\textwidth}
        \centering
        \includegraphics[width=\textwidth]{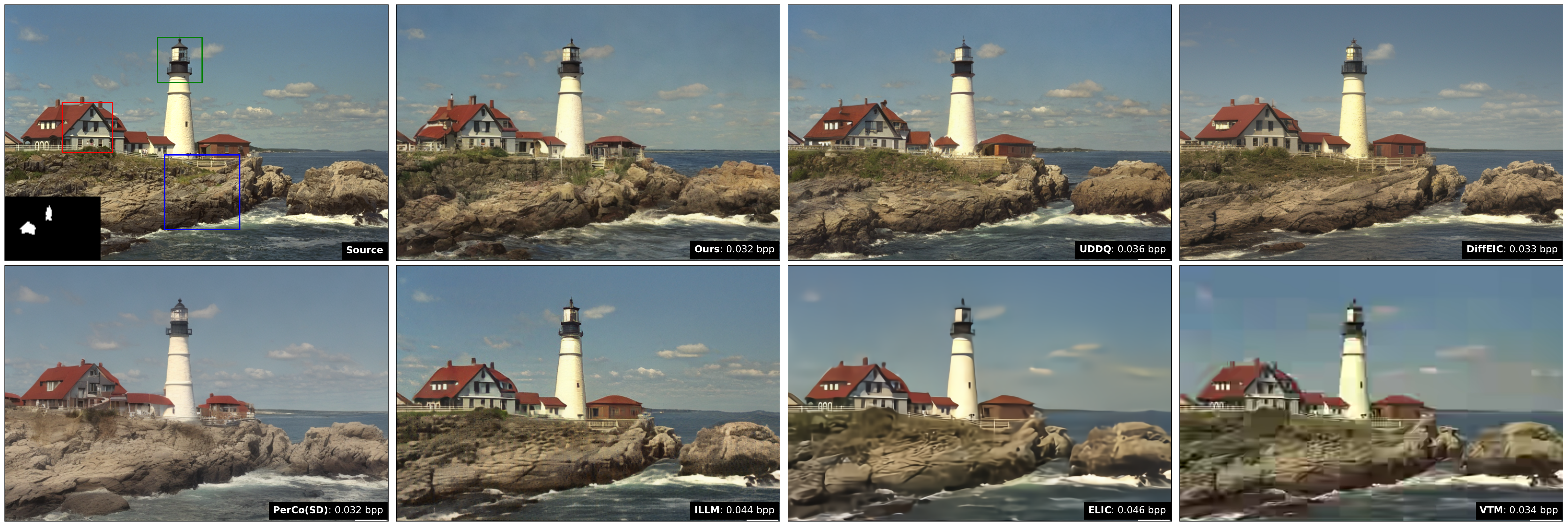}
    \end{subfigure}
    \begin{subfigure}{\textwidth}
        \centering
        \includegraphics[width=\textwidth]{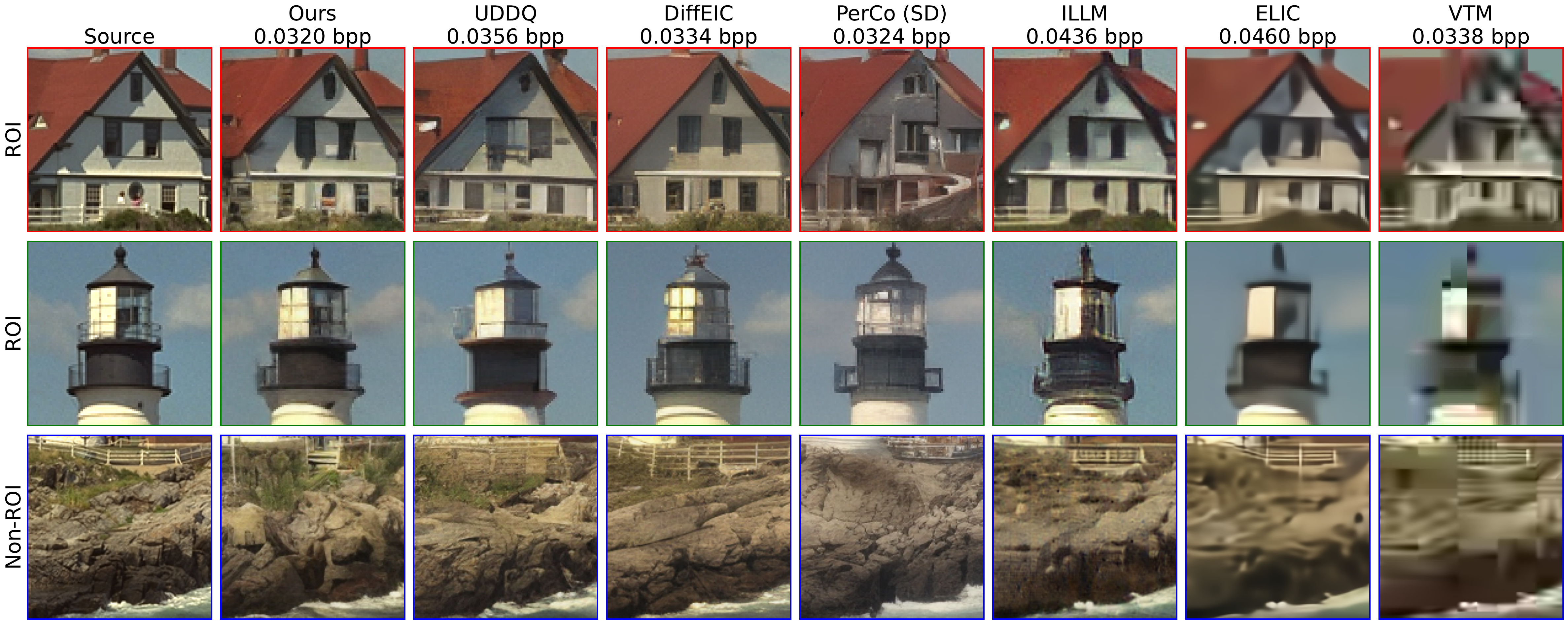}
    \end{subfigure}
    \caption{Full image (top) and zoomed-view (bottom) qualitative comparison between our method and the baselines. When considering the entire image, our method looks consistently realistic. In ROI areas, the reconstruction is faithful to the source, and areas outside of ROI can vary semantically but still remain plausible. Best viewed digitally.}
\label{fig:kodim21_full_and_crops}
\end{figure*}

Additional reconstructions, both full image and zoomed-in crops, are shown in \cref{fig:kodim06_full_and_crops} and \cref{fig:kodim21_full_and_crops}.
When considering the appearance of the entire image, our method is consistently realistic, with plausible textures and minimal artifacting.
Objects within an ROI are reconstructed accurately, closely matching the source image.
Outside of an ROI, image content can expectedly vary semantically, yet it remains plausible and consistent with neighboring areas.

\subsection{Quantitative Results}
\begin{figure*}
    \centering
    \includegraphics[width=\linewidth]{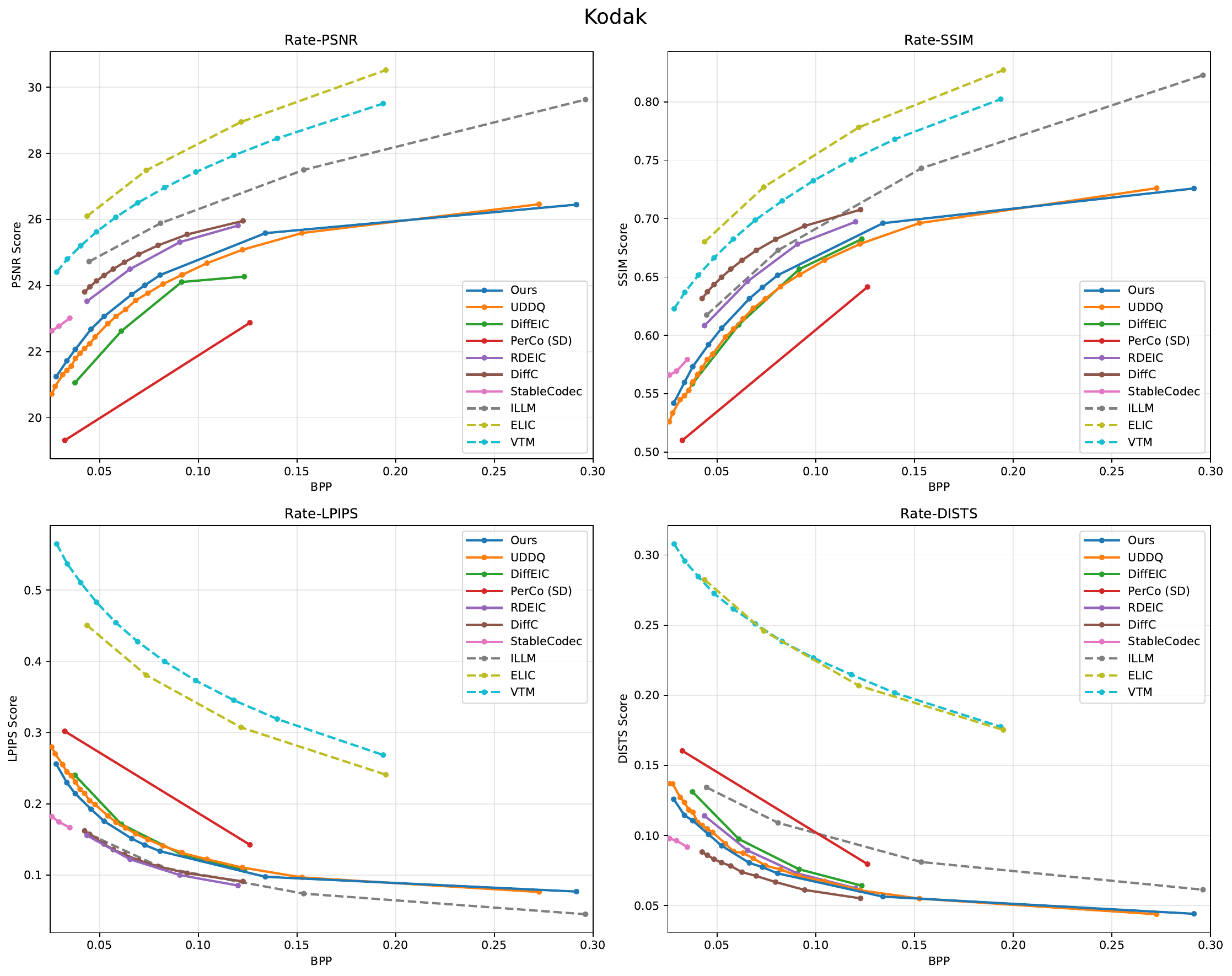}
    \caption{Quantitative rate-distortion comparison on the Kodak dataset. We measure with both pixelwise error metrics~(PSNR and SSIM, top row) and more perceptually oriented metrics~(LPIPS and DISTS, bottom row). Diffusion-based methods are shown with solid lines, while non-diffusion methods have dashed lines.}
    \label{fig:rd_full_kodak}
\end{figure*}

\begin{figure*}
    \centering
    \includegraphics[width=\linewidth]{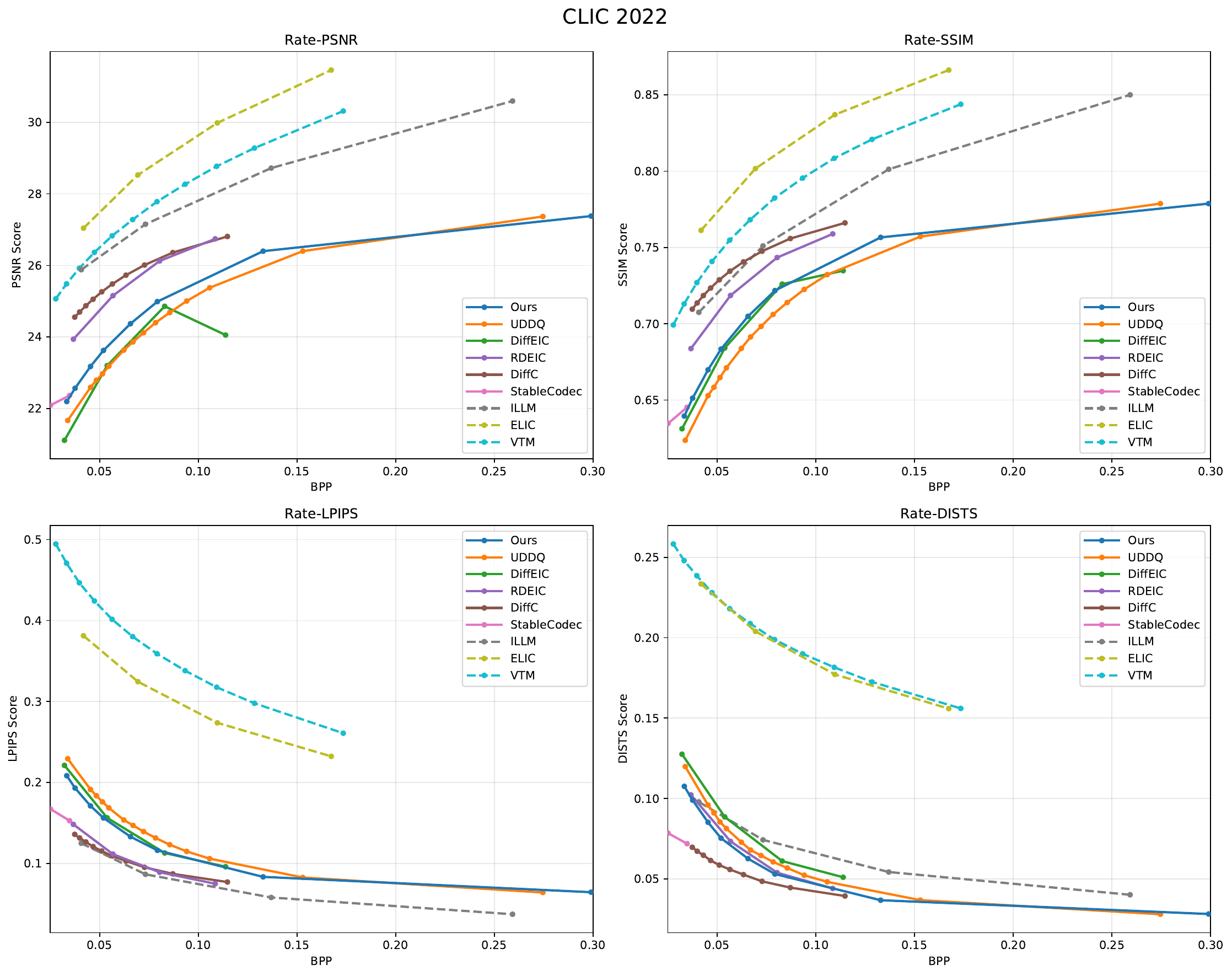}
    \caption{Quantitative rate-distortion comparison on the CLIC 2022 dataset. We measure with both pixelwise error metrics~(PSNR and SSIM, top row) and more perceptually oriented metrics~(LPIPS and DISTS, bottom row). Diffusion-based methods are shown with solid lines, while non-diffusion methods have dashed lines.}
    \label{fig:rd_full_clic22}
\end{figure*}

We show quantitative rate-distortion results with non-spatially adaptive generative baselines using PSNR, SSIM, LPIPS, and DISTS in \cref{fig:rd_full_kodak} and \cref{fig:rd_full_clic22}.
Our method uses a constant ROI map~(\ie, all image regions weighted evenly) in these experiments, as the other baseline methods do not support region adaptivity.
We also include additional generative baselines \textbf{RDIEC}~\cite{rdeic}, \textbf{StableCodec}~\cite{stablecodec} and \textbf{DiffC}~\cite{diffc}.

When considering realism and end user perceptual experience, as we do in this work, quantifying image quality using pixelwise error metrics such as PSNR or SSIM is proven and experimentally verified to be ineffective~\cite{blau2019Rethinking, mentzer2020HighFidelitya, relic2024Lossya}.
Metrics that measure distortion in the feature space of a deep convolutional model, for example LPIPS and DISTS, provide a more perceptually-oriented image quality metric, and are commonly used in evaluation of generative lossy compression methods~\cite{mentzer2020HighFidelitya, relic2024Lossya, careil2024image, li2025Extreme, relic2025Bridging, xu2025Decouplea}.
We include results on all above metrics for completeness and to provide further evidence of this misalignment.

As our goal is to add spatial adaptivity to generative codecs and not to achieve SoTA rate-distortion results in the spatially uniform case, our codec is not best performing in these metrics.
However, we remain competitive with these baselines in LPIPS and DISTS scores.
When considering pixelwise accuracy measured in PSNR and LPIPS, our method expectedly suffers compared to traditional and VAE-based codecs, as do the other diffusion-based methods.

\subsection{Ablation Study}
The ablation study in the main text includes an experiment where our spatially varying diffusion model is replaced with a spatially uniform diffusion model and modified RePaint~\cite{lugmayr2022RePainta} sampling.
Below we include specific implementation details of this experiment.

RePaint is originally a diffusion-based inpainting method, however we extend the general algorithm to work for our region-adaptive compression task.
In inpainting, there are two regions in the image, the unknown area, which is inpainted, and the known region, which is the context that remains the same.
The unknown regions are fully generated and thus start from pure noise (\ie, \(t=T\)) and iterate through \(t\in\{T, T-1, T-2, ..., 1\}\).
In the known regions, we have the ground truth image at \(t=0\).
Spatially uniform diffusion models require the entire image to be at the same \(t\), and RePaint presents a method to unify noise levels at every diffusion forward pass.
At each timestep \(t\), the known regions undergo the forward diffusion process \(q(\mathbf{y}_t|\mathbf{y}_0)\) to reach the same noise level as the unknown regions.
The diffusion model then denoises the entire image to \(t-1\), and afterwards the known regions are replaced with the ground truth pixels using the inpainting mask.\footnote{Fig. 2 in the original text \cite{lugmayr2022RePainta} visualizes this algorithm.}
This process repeats to fully generate the unknown regions using the known regions as context.

We modify this algorithm for our purpose by extending it from two regions at timesteps \(t=t\) and \(t=0\), to any number of regions at any arbitrary timestep.
At every \(t\in\{T, T-1, T-2, ..., 1\}\) the less noisy regions at timesteps \(s_1, s_2, ... < t\)\footnote{\(\{s_1, s_2, ...\}\) is the set of unique integer timesteps in \(\bt\).} are noised according to the forward process \(q(\mathbf{y}_t|\mathbf{y}_{s_i})\).
The image, now fully at timestep \(t\), is denoised with the diffusion model, and the less noisy regions are replaced with their known context.
In other words, we perform multi region inpainting using partially noisy context.
We use RePaint parameters \(n=1\) and \(j=1\) as this results in the same number of forward passes through the diffusion model as our method.
This method also requires a spatially uniform diffusion model that operates with uniform noise, which we implement according to Relic~\etal~\cite{relic2025Bridging}.

\subsection{UDDQ-ROI}
\begin{figure}[t]
    \centering
    \begin{subfigure}{\linewidth}
    \centering
        \includegraphics[width=\linewidth]{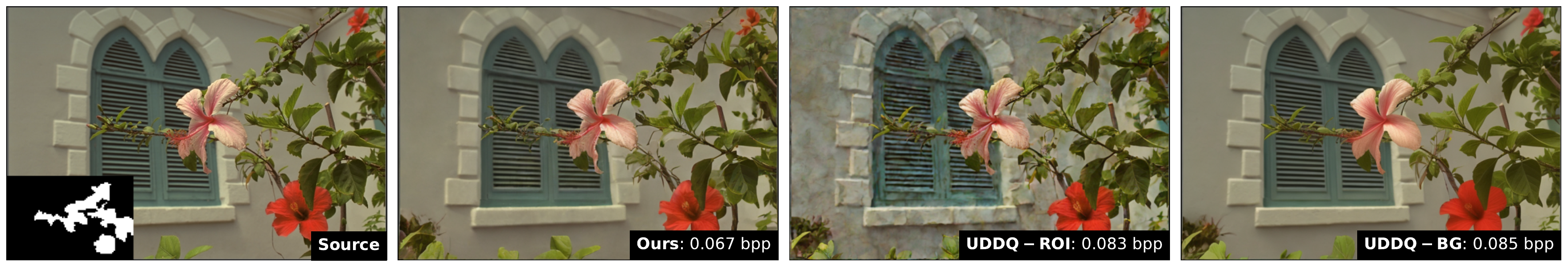}
    \end{subfigure}
    \begin{subfigure}{\linewidth}
    \centering
        \includegraphics[width=\linewidth]{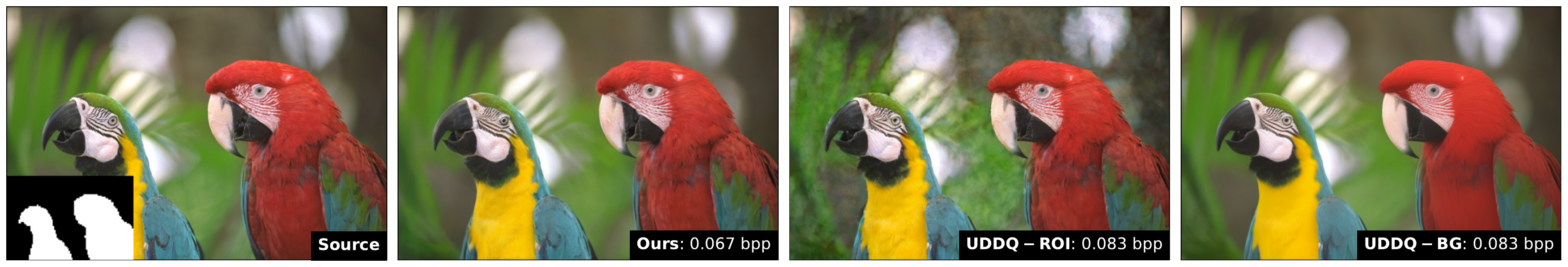}
    \end{subfigure}
    \caption{Visual examples of variants of our baseline comparison UDDQ-ROI and UDDQ-BG on the Kodak dataset. UDDQ-ROI reconstructs the ROIs accurately, but introduces textural artifacts in background regions. UDDQ-BG reconstructs the background well, but ROI areas are oversmoothed and saturated. Our method reconstructs both ROI and background regions realistically and faithfully.}
    \label{fig:baseline_variants}
\end{figure}

We build and compare against a naive generative spatially-adaptive baseline codec UDDQ-ROI in the main text.
We take the method presented in Relic~\etal~\cite{relic2025Bridging} and apply their proposed quantization scheme per-pixel, which allows for spatially adaptive rate control.
We then apply the number of diffusion decoding steps that reconstructs the ROIs with best quality.
This aligns with our goal of building a spatially adaptive generative baseline.
However, in principle it is possible to use any arbitrary number of diffusion steps.
One such example is to reconstruct the non-ROI areas with best quality (we refer to this variant as \textbf{UDDQ-BG}).
Although this may result in higher overall image quality, areas in ROI will have poorer perceptual quality, typically oversmoothed and saturated.
We visualize reconstructions of both baseline variants, as well as our method, in \cref{fig:baseline_variants}.

\section{Timestep Resampling}
\label{sec:timestep_resampling}
In this section we provide additional information regarding our timestep resampling process.
We first overview some preliminaries.
Our proposed codec follows the diffusion dequantization paradigm~\cite{relic2024Lossya, relic2025Bridging}, where the target bitrate and number of diffusion steps are closely related through the user input quantity \(\bt\).
It is easiest to think of \(\bt\) as the number of diffusion steps to run at the receiver; in fact, in the spatially uniform case, it is exactly that.
This defines the timestep in the diffusion process that the quantized latent \(\hat{\mathbf{y}}_{\bt}\) is at, and correspondingly, via \cref{eq:spatial_uniform_forward}, the noise level of \(\hat{\mathbf{y}}_{\bt}\).
We introduce this amount of noise by modulating the quantization bin width according to \cref{eq:our_forward}.
In the spatially uniform case, we simply denoise the entire image with \(\bt\) diffusion steps, which recovers the original image.

This presents issues, however, in the spatially varying case, as each different region must be denoised with a different number of diffusion steps.
Following this na\"ive strategy can affect the noise properties of diffusion~\cite{chang2023How} or inhibit interaction between neighboring regions~\cite{lugmayr2022RePainta}.
Our goal is therefore to gradually denoise all areas together, such that they reach the fully denoised state at the same diffusion iteration.
We do this by modulating the amount of noise removed at each diffusion step depending on its initial noise level, which we call timestep resampling.
Intuitively, using the example shown in \cref{fig:sampling}, the blue areas have ``4 steps worth of noise'' denoised in 7 diffusion iterations.
This is done as explained in \cref{subsec:spatial_diff}, where the \(\bt\) used to quantize in the forward process defines an initial \(\balpha_{\bt}\), and then, for each region, \(\text{max}(\bt)\) evenly spaced intervals are sampled between the initial \(\balpha_{\bt}\) and 1, the terminal \(\balpha_{\bt}\).

The observant reader may notice that we define and use a discrete time diffusion process, yet resample the timesteps at inference time, which may seem contradictory.
This is possible due to the nuances between DDPM sampling, used in training the diffusion model, and DDIM sampling, which we use at inference time.
When using DDPM sampling during training, typically a large \(T\) is used (in our case \(T=1000\), where \(t\in\{1000, 999, 998, ..., 1,0\}\).
DDIM sampling uses a subset of these original 1000 steps, for example \(t\in\{981, 961, 941, ...41, 21, 1\}\).
Our \(\bt\) values are defined with respect to the DDIM sampling indices~(\ie, \(\bt=4\) corresponds to \(t\in\{61,41,21,1\}\).
When performing our timestep resampling, we reindex the DDPM steps, therefore changing the denoising step size without rediscretizing the timesteps in our discrete process.
Using the example in \cref{fig:sampling}, the 4 timesteps of the blue region are \(t\in\{61,41,21,1\}\). \(\text{max}(\bt) =7\) (the yellow region), and thus we resample 7 DDPM timesteps starting from \(t=61\), yielding \(t\in\{61,51,41,31,21,11,1\}\).
In some cases it is possible for this reindexing to not align perfectly, however we do not observe any negative effects in practice.

\section{Arbitrary Mask Types}
Our method is capable of using any arbitrary ROI map inputs at inference time.
Below we show examples of reconstructions with various types of masks.
We include specific \(t\) values of each region in these examples, for a detailed explanation of these values see \cref{sec:timestep_resampling}.

\begin{figure*}
    \centering
    \includegraphics[width=\linewidth]{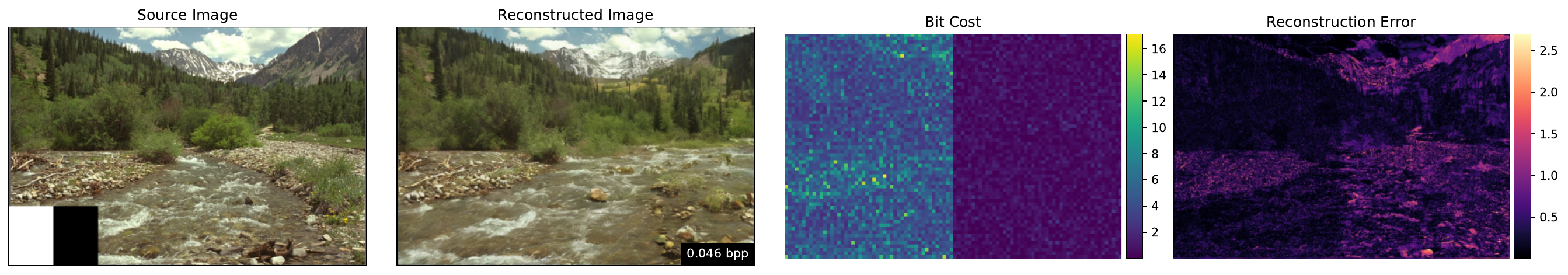}
    \caption{Bitrate allocation and reconstruction error of our method, where the left half of the image is compressed to high rate, and the right half to low rate. Bit cost is visualized in the third column. The right half of the reconstructed image (second column) is semantically different, yet still plausible. Best viewed zoomed.}
    \label{fig:rate_allocation}
\end{figure*}

\cref{fig:rate_allocation} shows a reconstruction where the left half of the image is compressed to high rate (\(t=5, ~0.0753\text{ bpp}\)), and the right half to extremely low rate (\(t=50, ~0.0099\text{ bpp}\)).
Correspondingly, the reconstruction error of the right side is higher, although critically that region of the reconstructed image is still realistic and semantically plausible.
This spatial bit allocation and error map can be seen in the third and fourth columns, respectively.

\begin{figure*}
    \centering
    \includegraphics[width=\linewidth]{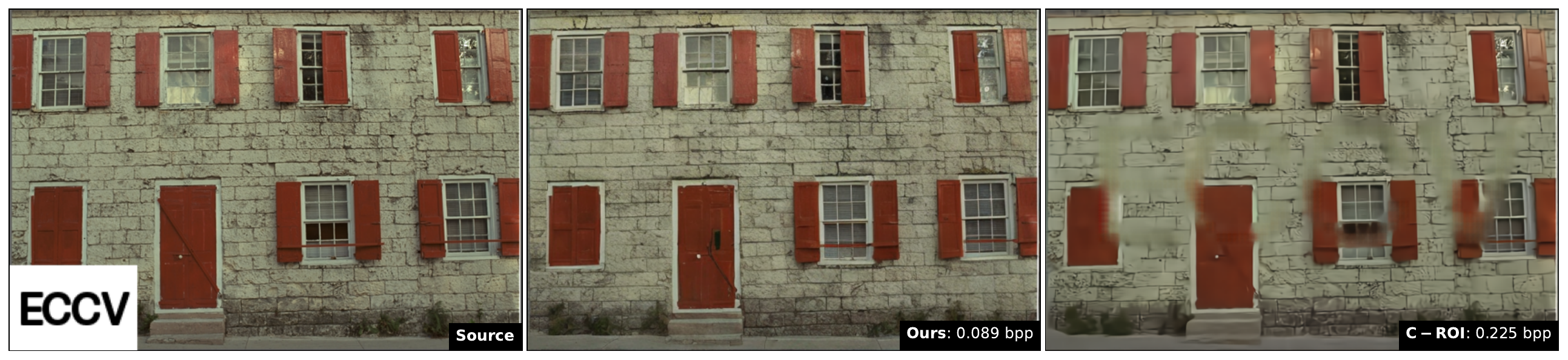}
    \caption{Visualization of the reconstruction using a complex mask. Our approach does (middle) not suffer from boundary artifacts in the masked region, which is encoded at a low rate, and the content is realistically reconstructed using neighboring context. Previous methods (right), contain noticeable artifacts.}
    \label{fig:text_mask}
\end{figure*}

The image in \cref{fig:text_mask} was compressed with an ROI mask that contains complex boundaries.
The ``ECCV'' text is compressed to low rate (\(t=40\)), while the rest of the image to high rate (\(t=5\)).
Our method's reconstruction contains no boundary artifacts, and it is very difficult to tell where the mask boundaries lie when only looking at the reconstructed image.
Furthermore, the text region compressed to low rates is realistically reconstructed using the context from neighboring regions, particularly visible when comparing our reconstruction to that C-ROI~\cite{jin2025Customizable}.

\begin{figure*}
    \centering
    \includegraphics[width=\linewidth]{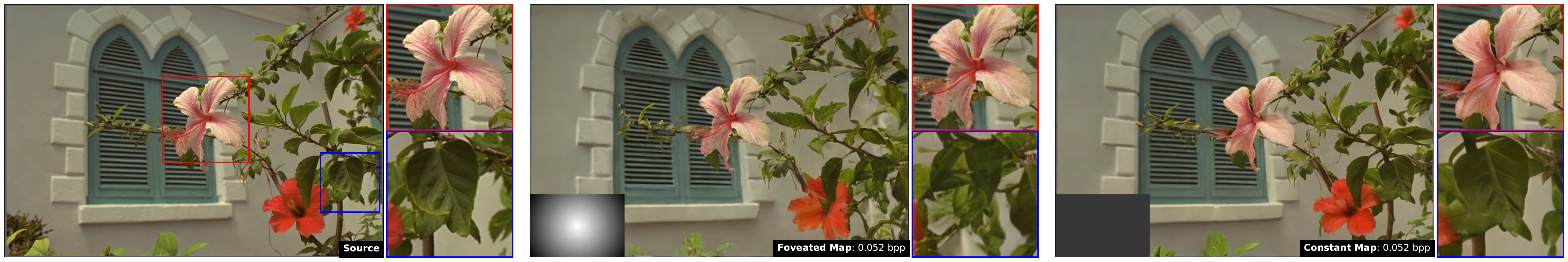}
    \caption{Reconstruction of our method using a ``foveated'' mask, aiming to exploit human foveal vision. The center of the image (flower crop) is most accurately reconstructed, and visual quality is gradually reduced towards the edges (leaf crop), which saves bits while being less noticeable.}
    \label{fig:foveated_mask}
\end{figure*}

We also experimented with a ``foveated'' mask, which aims to exploit human foveal vision, where the center of vision has highest visual acquity that gradually decreases towards the periphery.
To this end, we use an ROI map that allocates most bits to the center of the image, and gradually less towards the image edges, shown in \cref{fig:foveated_mask}.
The flower, located at the center of the image, is reconstructed most accurately.
Image quality decreases towards the edges, such as the leaves on the right side, however correspondingly less bits are spent in those areas.
This complex mask highlights the inherent flexibilty of our method to use any ROI mask with multiple regions at multiple different target bitrates.
Additionally, we feel this shows an interesting application of our method to future downstream tasks, such as compression of content for virtual reality headsets.

\section{User Study}
\begin{figure*}
    \centering
    \includegraphics[width=\linewidth]{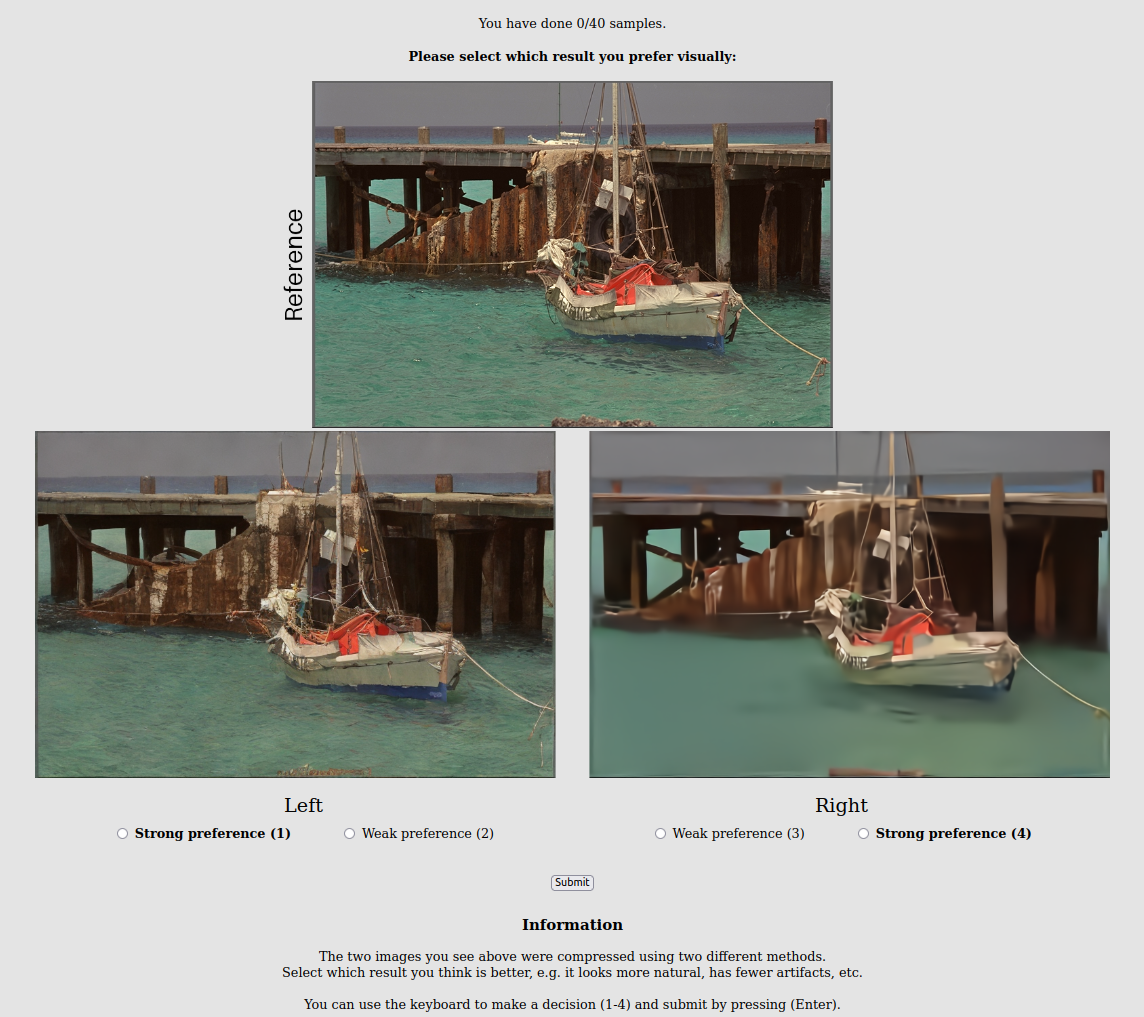}
    \caption{The interface of our user study. Participants are shown reconstructions of two methods, as well as the ground truth, and asked to choose their preference.}
    \label{fig:userstudy_ui}
\end{figure*}

We adopt the 2-alternative forced choice (2AFC) strategy for our user study.
Participants are presented reconstructions from 2 different methods, placed side by side, and a ground truth reference centered above.
A screenshot of our interface is shown in \cref{fig:userstudy_ui}.
Participants are asked to choose which reconstruction they prefer, choosing between ``strong preference'' and ``weak preference'' for either image.
There is no option to skip or choose equal preference.

The study was conducted using reconstructions on a subset of the Kodak dataset.
Comparison sets were selected as follows:
given all possible comparison sets~(\ie every subset of one image per method from all reconstructions at all bitrates of all baseline methods), we filtered by choosing those only where our method had the lowest bitrate among the set.
This yielded 13 out of a possible 24 image samples from the original dataset.
We then ranked valid comparison sets by ``coefficient of variation'', which we defined as the standard deviation of bitrates within the comparison divided by the mean bitrate (to avoid bias towards lower rates).
For each image sample we selected the comparison set with the lowest coefficient of variation.
This yielded an image set where the bitrates between \emph{all} methods are closest, but not that our method's reconstruction has the closest rate to each of the baselines.
Therefore, for each baseline reconstruction from the identified best comparison set, we chose the reconstruction from our method with the closest, yet still lower, bitrate.

We collected 745 votes from 16 participants, an average of 46.6 ratings per participant.
Each participant was asked to complete 40 ratings, however, could continue voting until all possible comparisons were exhausted (a maximum of 91).
Our sampler was set up such that each comparison has a roughly equal number of votes.

\section{Reproduction of Baselines}
Below we detail specific information on how reconstructions were obtained from the baselines for comparison.

\paragraph{C-ROI.~\cite{jin2025Customizable}} We produce reconstructions using the official code release and model weights, available at \url{https://github.com/hccavgcyv/Customizable-ROI-Based-Deep-Image-Compression} (commit \texttt{fea0dfa}).
We use the \(\sigma=0.01\) models, and modify the code to use our ROI maps instead of text-based ROIs.

\paragraph{Relic~\etal~(UDDQ).~\cite{relic2025Bridging}}
Reconstructions were obtained through personal communication.

\paragraph{DiffEIC.~\cite{li2025Extreme}}
We produce reconstructions using the official code and released model weights, available at \url{https://github.com/huai-chang/DiffEIC} (commit \texttt{c689ca7}).

\paragraph{RDEIC.~\cite{rdeic}}
We produce reconstructions using the official code and released model weights, available at \url{https://github.com/huai-chang/RDEIC} (commit \texttt{341ba4b}).

\paragraph{PerCo.~\cite{careil2024image}}
No official implementation of PerCo is publicly available, and cannot be reproduced due to their use of proprietary latent diffusion model.
We instead use PerCo (SD)~\cite{korber2024PerCo}, a third-party implementation using Stable Diffusion v2.1 as the diffusion backbone.
Code and model weights are available at \url{https://github.com/Nikolai10/PerCo} (commit \texttt{da3be39}).

\paragraph{StableCodec~\cite{stablecodec}}
We produce reconstructions using the official code and released model weights, available at \url{https://github.com/LuizScarlet/StableCodec} (commit \texttt{e7994cf}).

\paragraph{DiffC.~\cite{diffc}}
We produce reconstructions using the official code and released model weights, available at \url{https://github.com/JeremyIV/diffc} (commit \texttt{e6771e0}).

\paragraph{ILLM.~\cite{muckley2023Improving}}
Reconstructions are produced using the official code and model weights, available at \url{https://github.com/facebookresearch/NeuralCompression/tree/main/projects/illm} (commit \texttt{47438f9}).

\paragraph{ELIC.~\cite{he2022ELICa}}
As no official code is available, we produce reconstructions using a popular third-party reimplementation, available at \url{https://github.com/VincentChandelier/ELiC-ReImplemetation} (commit \texttt{92a9ece}).
At low bitrates, rate-distortion performance of this implemenation is nearly identical to the official reported metrics.

\paragraph{VTM.~\cite{jointvideoexpertsteamjvet2020VVC}}
We use the official H.266 reference implementation VTM, version 22.2, available at \url{https://vcgit.hhi.fraunhofer.de/jvet/VVCSoftware_VTM} (tag \texttt{VTM-22.2}).
RGB source images are converted to and from YCbCr 4:4:4 format using ITU-R BT.709 coefficients before and after coding, respectively.
We then run the following command:
\begin{verbatim}
EncoderAppStatic \
-c encoder_intra_vtm.cfg \
--InputFile=[INPUT] \
--BitstreamFile=[BITSTREAM] \
--ReconFile=[OUTPUT] \
--SourceWidth=[WIDTH] \
--SourceHeight=[HEIGHT] \
--InputBitDepth=8 \
--OutputBitDepth=8 \
--OutputBitDepthC=8 \
--InputChromaFormat=444 \
--FrameRate=1 \
--FramesToBeEncoded=1 \
--QP=[QP] \
--ConformanceWindowMode=1 \
--InputColourSpaceConvert=RGBtoGBR \
--SNRInternalColourSpace=1 \
--OutputInternalColourSpace=0 \
\end{verbatim}

\end{document}